%% file: main.tex
\documentclass[a4paper,11pt,amsmath,amssymb]{article} 
\usepackage{jheppub}
\usepackage{amsmath} 
\usepackage{array,relsize,float}
\usepackage{pstricks}
\usepackage{color} 
\usepackage{amssymb}
\usepackage{slashed}
\usepackage{tikz-cd}
\usepackage{graphicx} 
\usepackage{epsfig}
\usepackage{multicol} 
 \usepackage{xspace}
 \usepackage[T1]{fontenc}
 
\usepackage{slashed} 
\usepackage{hyperref}
\usepackage{pdfpages}
\usepackage{array}
\usepackage{tikz-cd}
\usepackage{amsfonts,layout,appendix,subfigure}

\allowdisplaybreaks
\input{epsf}

\newcommand{\nn}{\nonumber}

\renewcommand{\ln}[1]{\log \left(#1\right)}
\newcommand{\uv}{{\rm UV}}

\newcommand{\td}[1]{\tilde{\delta}\left(#1\right)}
\newcommand{\ii}{\imath 0}

\newcommand{\qon}[1]{q_{#1,0}^{(+)}}

\def\beq{\begin{equation}} 
\def\eeq{\end{equation}} 
\def\beqn{\begin{eqnarray}} 
\def\eeqn{\end{eqnarray}} 
 
\def\to{\rightarrow}
 
\def\nn{\nonumber}

\def\Mathematica{{{\sc Mathematica}}}
\def\Fiesta{{{\sc Fiesta 4.2}}}
\def\SecDec{{{\sc SecDec 3.0}}}
\def\FiniteFlow{{{\sc FiniteFlow}}}

\renewcommand{\thefigure}{\arabic{figure}}

\def\beq{\begin{equation}} \def\eeq{\end{equation}}
\def\beqn{\begin{eqnarray}} \def\eeqn{\end{eqnarray}}
 \def\to{\rightarrow}
\def\nn{\nonumber}

\def\ln#1{\mathrm{log}\left(#1\right)}
\def\Eq#1{Eq.~(\ref{#1})}
\def\qon#1{q_{#1,0}^{(+)}}

\def\beq{\begin{equation}}
\def\eeq{\end{equation}}
\def\bea{\begin{eqnarray}}
\def\eea{\end{eqnarray}}
\def\beqn{\begin{eqnarray}} \def\eeqn{\end{eqnarray}}
\def\beeq{\begin{eqnarray}}
\def\eeeq{\end{eqnarray}}
\newcommand{\be}{\begin{equation}}
\newcommand{\ee}{\end{equation}}

\def\nn{\nonumber}
\def\Eq#1{Eq.~(\ref{#1})}
\def\ln#1{\mathrm{log}\left(#1\right)}

\def\td#1{\tilde{\delta}\left(#1\right)}

\newcommand\g{g_{\mathrm{S}}}

\def\fun#1#2{\lower3.6pt\vbox{\baselineskip0pt\lineskip.9pt
\ialign{$\mathsurround=0pt#1\hfil ##\hfil$\crcr#2\crcr\sim\crcr}}}

\newcommand{\mcal}[1]{\mathcal{#1}}

\newcommand{\eq}[1]{Eq.~(\ref{#1})}
\def\qon#1{q_{#1,0}^{(+)}}

\newcommand{\iz}{\imath0}

\newcommand{\LL}{\left}
\newcommand{\RR}{\right}

\def\uv{{\rm UV}}

\def\ii{\imath 0}

\newcommand\numberthis{\addtocounter{equation}{1}\tag{\theequation}}

\pdfoutput=1

\newcommand{\valencia}{Instituto de F\'{\i}sica Corpuscular, Universitat de Val\`{e}ncia -- Consejo Superior de Investigaciones Cient\'{\i}ficas, Parc Cient\'{\i}fic, E-46980 Paterna, Valencia, Spain.}
\newcommand{\culiacan}{Facultad de Ciencias F\'isico-Matem\'aticas, Universidad Aut\'onoma de Sinaloa, 
    Ciudad Universitaria, CP 80000 Culiac\'an, Mexico.}
\newcommand{\culiacana}{Facultad de Ciencias de la Tierra y el Espacio, Universidad Aut\'onoma de Sinaloa, Ciudad Universitaria, CP 80000 Culiac\'an, Mexico.}
\newcommand{\berlin}{Deutsches Elektronen-Synchrotron, DESY, Platanenallee 6, D–15738 Zeuthen, Germany.}
\newcommand{\munich}{Max-Planck-Institut f{\"u}r Physik, Werner-Heisenberg-Institut, 80805 M{\"u}nchen, Germany.}
\newcommand{\napoli}{Dipartimento di Fisica, Universit\`{a} di Napoli Federico II and
INFN, Sezione di Napoli, I-80126 Napoli, Italy.}

\begin{document}

\title{A stroll through the loop-tree duality}
\author[a]{Jes\'us Aguilera-Verdugo,}
\author[a]{F\'elix Driencourt-Mangin,}
\author[b]{Roger J. Hern\'andez-Pinto,}
\author[a]{Judith Plenter,}
\author[c]{Renato Maria Prisco,}
\author[a,d]{Selomit Ram\'irez-Uribe,}
\author[a]{Andr\'es Renter\'ia-Olivo,}
\author[a]{Germ\'an Rodrigo,}
\author[e,a]{German Sborlini,}
\author[f]{William J. Torres Bobadilla,}
\author[c]{Francesco Tramontano}
\affiliation[a]{\valencia}
\affiliation[b]{\culiacan}
\affiliation[c]{\culiacana}
\affiliation[d]{\napoli}
\affiliation[e]{\berlin}
\affiliation[f]{\munich}

\emailAdd{german.rodrigo@csic.es, roger@uas.mx, german.sborlini@desy.de}

\preprint{IFIC/21-11, DESY 21-056, MPP-2021-65}

\abstract{The Loop-Tree Duality (LTD) theorem is an innovative technique to deal with multi-loop scattering amplitudes, leading to integrand-level representations over an Euclidean space. In this article, we review the last developments concerning this framework, focusing on the manifestly causal representation of multi-loop Feynman integrals and scattering amplitudes, and the definition of dual local counter-terms to cancel infrared singularities.}

\setcounter{page}{1}
\maketitle

\section{Introduction}
\label{sec:intro}
The most successful description of the microscopic structure of Nature is currently given by the Standard Model (SM), which is based on quantum field theories (QFT) with specific gauge symmetries. The strong interplay between experiment and theory relies on confronting precise predictions with accurate data. In the context of high-energy particle physics, the Large Hadron Collider (LHC) and the planned future colliders \cite{Abada:2019lih,Abada:2019zxq,Benedikt:2018csr,Abada:2019ono,Blondel:2019vdq,Bambade:2019fyw,Roloff:2018dqu,CEPCStudyGroup:2018ghi} will keep on reducing the experimental uncertainties, forcing to reach a compatible accuracy level from the theory side. However, there are huge bottlenecks that prevent straightforward calculations.

Due to their complexity, exact solutions of QFTs are not available for generic scattering processes. Thus, extracting predictions from theory requires to use approximations, whose applicability is restricted to specific situations. Whilst lattice methods are reliable in the low-energy regime, perturbation theory is the most suitable strategy to tackle the phenomenological description of particle collisions at high-energies. Even if the perturbative approach reduces the original calculation to a series expansion in the interaction couplings, adding higher-orders is far from trivial due to the presence of complicated phase-space and Feynman loop integrals. Moreover, these objects feature numerical instabilities originated in threshold configurations, as well as other singularities that cancel only after putting all the contributions together.

In the last forty years, a huge effort has been done to develop new methodologies for an efficient calculation of physical observables at higher-orders. As a first step, proper regularization methods have been applied in order to make explicit the singular structures of these objects. One of the customary choices is Dimensional Regularization (DREG), which consists in modifying the number of space-time dimensions to achieve integrability \cite{tHooft:1972tcz,Bollini:1972ui,Cicuta:1972jf,Ashmore:1972uj}. However, in the context of QFT, changing the number of space-time dimensions leads to definition problems, such as the $\gamma_5$ \cite{Gnendiger:2017rfh,Bruque:2018bmy}, and prevents a straightforward numerical implementation. For this and other technical reasons, there is an ongoing effort in the high-energy physics community to develop new strategies that locally regularise QFT whilst keeping the standard four dimensions of the space-time \cite{Pittau:2012zd,Fazio:2014xea,Mastrolia:2015maa,Primo:2016omk,Gnendiger:2017pys,TorresBobadilla:2020ekr}.

Once the Feynman amplitudes and phase-space integrals are properly regularised, we need to compute them and combine all the ingredients. On the one hand, this requires to calculate multi-loop multi-leg Feynman integrals. Several techniques are currently available for this purpose: Feynman parametrisation, Mellin-Barnes transformations, IBP identities \cite{Chetyrkin:1981qh,Laporta:2001dd}, sector decomposition~\cite{Binoth:2000ps,Smirnov:2008py,Carter:2010hi,Borowka:2017idc}, semi-numerical approaches~\cite{Francesco:2019yqt,Bonciani:2019jyb,Czakon:2008zk}, among others \cite{Heinrich:2020ybq}. Very recently, algebraic geometry \cite{Larsen:2015ped,Bern:2017gdk,Zeng:2017ipr,Boehm:2017wjc,Boehm:2018fpv,Bendle:2019csk} was used to develop multi-loop integrand reduction algorithms~\cite{Mastrolia:2011pr,Badger:2012dp,Zhang:2012ce,Mastrolia:2012an,Mastrolia:2012wf,Ita:2015tya,Mastrolia:2016dhn,Ossola:2006us}, and to explore alternative representations of Feynman integrals \cite{Baikov:1996rk,Frellesvig:2017aai}. Also, unitarity based methods \cite{Peraro:2016wsq,Badger:2017jhb,Abreu:2017hqn,Badger:2019djh} and intersection theory \cite{Mastrolia:2018uzb,Frellesvig:2019kgj,Frellesvig:2019uqt,Weinzierl:2020xyy} are being studied to improve the reduction to master integrals and their subsequent evaluation. 

On the other hand, loop contributions must be combined with real-emission corrections and suitable counter-terms to cancel the infrared (IR) and ultraviolet (UV) singularities, leading to a reliable prediction of the physical cross-sections and differential distributions. There are several strategies to achieve this cancellation, most of them based on the subtraction methods \cite{Kunszt:1992tn,Frixione:1995ms,Catani:1996jh}. Generally speaking, these frameworks involve adding and subtracting counter-terms which locally cancel the IR behaviour of the real part, but at the same time they are easy enough to be integrated analytically and subtracted from the virtual contribution. There are several variations of this approach: dipole subtraction \cite{Catani:1996jh,Catani:1996vz}, $q_T$ subtraction/resummation \cite{Catani:2000vq,Catani:2013tia}, antenna subtraction \cite{GehrmannDeRidder:2005cm}, colorful subtraction \cite{Kardos:2016qol,DelDuca:2016ily}, among others. Also, there are alternatives that rely on a different way to achieve the cancellation of singularities, like the n-jetiness \cite{Gaunt:2015pea,Boughezal:2015dra} or the local analytic subtraction frameworks \cite{Magnea:2018hab,Magnea:2019qfs}.

Besides those methods, a novel strategy to tackle, simultaneously, the efficient calculation of loop integrals and physical observables was developed. This framework is based on the Loop-Tree Duality (LTD) theorem \cite{Catani:2008xa,Rodrigo:2008fp,Bierenbaum:2010cy,Bierenbaum:2012th}, which opens loops into trees and recasts virtual states into configurations that resemble real-radiation processes. From the mathematical point of view, the LTD transforms the integration domain of loop integrals into an Euclidean space. In this way, a more intuitive understanding of the regions responsible for the singular structure of the loop integrals emerges \cite{Buchta:2014dfa}. This knowledge can be used to explore novel techniques pointing towards more efficient numerical implementations \cite{Buchta:2015xda,Buchta:2015wna}, integrand-level simplifications through asymptotic expansions \cite{Driencourt-Mangin:2017gop,Plenter:2019jyj,Plenter:2020lop} and unveiling the universal structures of scattering amplitudes at higher-orders \cite{Jurado:2017xut,Driencourt-Mangin:2019aix,Driencourt-Mangin:2019sfl,Driencourt-Mangin:2019yhu}. Moreover, the LTD provides the perfect framework to handle cross-section calculations in a fully unified way. Since the dual representation of loop integrals is defined in Euclidean domains, the virtual and real contributions can be directly combined at integrand-level to achieve a fully local cancellation of IR singularities. This is the so-called \emph{Four-Dimensional Unsubtraction} (FDU) approach \cite{Hernandez-Pinto:2015ysa,Sborlini:2016fcj,Sborlini:2016gbr,Sborlini:2016hat}, that profits from a local cancellation of singularities which makes it possible to bypass additional regulators (such as DREG). Also, this formalism allow us to write local UV counter-terms that exactly reproduce the expected results in conventional renormalisation schemes \cite{Sborlini:2016gbr,Sborlini:2016hat,Driencourt-Mangin:2019sfl,Driencourt-Mangin:2019yhu}, as well as fully local IR dual counter-terms \cite{Prisco:2020kyb}.

During the last two years, we have explored new features of the LTD approach. In particular, we discovered that a more general analysis of the singular structure of multi-loop multi-leg amplitudes is possible. By inquiring into the physical and anomalous thresholds of one and two-loop amplitudes \cite{Aguilera-Verdugo:2019kbz}, we proved that it was possible to get rid of non-physical singularities and retain only those contributions compatible with causality. Several studies about the causal structure of scattering amplitudes are available in the literature, using different techniques \cite{Cutkosky:1960sp,Tomboulis:2017rvd,Runkel:2019yrs,Runkel:2019zbm}. In Ref.~\cite{Verdugo:2020kzh}, we presented for the first time a manifestly causal integrand-level representation inspired by the LTD theorem. This was applied to remove unphysical threshold singularities and obtain causal integrand-level representations of several topological families of multi-loop multi-leg Feynman integrals \cite{Capatti:2019edf,Capatti:2019ypt,Verdugo:2020kzh,Ramirez-Uribe:2020hes,Aguilera-Verdugo:2020kzc,Aguilera-Verdugo:2020nrp,MANIFESTLYCAUSAL,Capatti:2020ytd}. Moreover, all-order causal formulae were obtained using novel algebraic relations \cite{Bobadilla:2021rmu}, leading to an automatised framework to implement these calculations \cite{TorresBobadilla:2021dkq}. A purely geometrical interpretation \cite{Sborlini:2021owe} was also developed, showing a complementary approach to tackle the causal structure of multi-loop multi-leg scattering amplitudes and Feynman integrals.

The purpose of this review is to give a brief summary of the latest developments regarding the LTD-based methods. To ease the presentation, it is organised in two parts. In the first one, starting in Sec. \ref{sec:PaperPRL}, we settle the notation of the LTD formalism. Special emphasis is put on the manifestly causal representation of loop integrands in the Euclidean space, as well as its derivation from Cauchy's residue theorem. The basis for a rigorous mathematical treatment of the nested residue strategy is given in Sec. \ref{sec:PaperJesus}. Then, we discuss the advantages of the Euclidean representation of loop integrands. In Sec. \ref{sec:PaperJudith}, we focus on the asymptotic expansion, whilst we discuss how to obtain compact causal formulae for generic topological families of multi-loop multi-leg Feynman integrals and scattering amplitudes in Secs. \ref{sec:PaperWilliam} and \ref{sec:PaperSelo}. In Sec. \ref{sec:CausalityNEW}, a few words are given to explain novel developments suggesting a deep connection among graph theory, algebra and the causal structure of multi-loop scattering amplitudes.

After that, we start the second part of the review, where we center into the computation of physical observables exploiting the LTD approach. In Sec. \ref{sec:FDUinto} we discuss the basis of the FDU formalism, showing a physical examples. Then, in Sec. \ref{sec:PaperFrancesco}, the construction of local counter-terms for cancelling IR singularities is presented with more general examples, including $\gamma^* \to 3\, {\rm jets}$ at NLO. Finally, we present the conclusions and future research directions in Sec. \ref{sec:Outlook}.

\section{Causality within the LTD formulation}
\label{sec:PaperPRL}
\input{draft_Andres}

\section{Mathematical properties of the nested residues}
\label{sec:PaperJesus}
\input{draft_JJ}

\section{Asymptotic expansions within LTD}
\label{sec:PaperJudith}
\input{draft_Judith}

\section{Manifestly causal representation and numerical efficiency}
\label{sec:PaperWilliam}
\input{draft_GW}

\section{Universal opening at four-loops}
\label{sec:PaperSelo}
\input{draft_Selo}

\section{Novel developments on causality}
\label{sec:CausalityNEW}
\input{draft_GS}

\section{Four-Dimensional Unsubtration (FDU)}
\label{sec:FDUinto}
\input{draft_Roger}

\section{Local dual counter-terms for cross-sections}
\label{sec:PaperFrancesco}
\input{draft_Francesco}

\section{Outlook and further developments}
\label{sec:Outlook}
In this manuscript, we reported on the recent developments in the numerical evaluation of multi-loop scattering amplitudes through the application of the Loop-Tree Duality (LTD) formalism. Based on the original approach, at one and two loops~\cite{Catani:2008xa,Bierenbaum:2010cy}, we presented a novel formulation of LTD~\cite{Verdugo:2020kzh}, which allowed to bootstrap the treatment of Feynman integrals, regardless of the loop order, and shed light to a complete automation. In effect, the decomposition of two- and three-loop scattering amplitudes in terms of dual integrands was obtained by introducing the concepts of maximal (MLT), next-to-maximal (NMLT) and next-to-next-to-maximal (N$^{2}$MLT) loop topologies. These families of diagrams share general properties that allow allows for the generalization of the dual representation at any loop order. 

Since LTD heavily relies on the application of the Cauchy residue theorem, we studied in details the treatment of multivariate rational functions in Ref.~\cite{Aguilera-Verdugo:2020nrp}, giving, in this way, a mathematical support to the novel formulation of Ref.~\cite{Verdugo:2020kzh}. Likewise, we proved all conjectures that were provided in the latter, thus, rendering in a well organised formalism ready to focus on physical applications. Hence, in view of the LTD decompositions of up-to three-loop scattering amplitudes, the natural extension, along the lines of Ref.~\cite{Verdugo:2020kzh}, was considering the four-loop case~\cite{Ramirez-Uribe:2020hes}. In effect, the complexity in the treatment of four-loop topologies increased because, differently from the two- and three-loop cases, this loop order was not described by only one loop topology, as in the lower cases. Hence, we introduced new configurations that appears at four loops: the N$^{3}$MLT and N$^{4}$MLT families. In this way, our formalism allowed a complete understanding and decomposition of the dual representation of any up-to four-loop scattering amplitude. 

In the spirit of profiting from the dual representation of integrands through LTD, we observed that the complete sum of the latter, for a given Feynman integral, led to a representation of an integrand that only manifests physical information. In other words, unphysical singularities or \emph{pseudo-thresholds} cancel out when all dual integrands are summed up. This is a remarkable feature of LTD that was originally observed in Refs.~\cite{Driencourt-Mangin:2019aix,Aguilera-Verdugo:2019kbz,Verdugo:2020kzh} and extensively studied in Ref.~\cite{Aguilera-Verdugo:2020kzc} for MLT, NMLT and N$^{2}$MLT configurations. In fact, the work performed in Ref.~\cite{Aguilera-Verdugo:2020kzc} made a comparison between the numerical stability of dual and causal integrands. While in the former numerical instabilities, originated from pseudo-thresholds, were present, the latter displayed a smooth behaviour: the numerical evaluation was much more stable and this allowed a straightforward computation of the Feynman integrals. In particular, we considered the numerical integration of ultraviolet finite integrals at three and four loops with presence of several kinematic variables.

The causal representation of scattering amplitudes has recently been a very active topic due to the simple structure of these integrands in terms of causal (physical) thresholds. In fact, an understanding of the structure of the latter started to appear in the literature by means of geometric properties~\cite{Sborlini:2021owe} and is yet under consideration. On top of the former approach, it was recently conjectured a breathtaking formulation of the all-loop causal representation of multi-loop scalar integrands, obtained from the features that characterised a topology, cusps and edges~\cite{Bobadilla:2021rmu}. This formulation provides, regardless of the loop order, the most symmetric causal representation of Feynman integrands, in a complementary and independent way to the nested residue calculation used by the LTD formalism. Thus, to give a support to these conjectures, we provide the \textsc{Mathematica} package \textsc{Lotty} to automate both dual and causal representation of scattering amplitudes~\cite{TorresBobadilla:2021dkq}. 

From the above-mentioned discussion, we claim that a treatment of Feynman integrands at an arbitrary number of loops is under control. Moreover, this is not the end of the story when comparing theoretical predictions with data derived by collider experiments. In effect, one yet has to deal with ultraviolet (UV) and infrared (IR) singularities. While the framework at next-to-leading order (NLO) is completely understood by means of the Four-Dimensional Unsubtraction scheme~\cite{Hernandez-Pinto:2015ysa,Sborlini:2016gbr,Sborlini:2016hat} and the dual subtractions \cite{Prisco:2020kyb}, a clear path towards calculations at higher orders (NNLO and beyond) needs to be devised. Preliminary studies of IR-safe scattering amplitudes has been recently considered in Refs.~\cite{Driencourt-Mangin:2019aix,Driencourt-Mangin:2019yhu}. Besides, the extension to the calculation of cross sections at NNLO is under study, in particular, focusing on a careful \emph{local} treatment of IR singularities. \\

We expect that the ideas presented in this manuscript will allow to elucidate in detail the state-of-the-art of the LTD formalism and the causal representation of scattering amplitudes by means of the latter. Likewise, the ideas that we have developed will certainly supply the scientific community with a powerful strategy to tackle the calculation of multi-loop Feynman integrals, in which phenomenological and formal applications can efficiently be carried out.

\section*{Acknowledgements}
This research was supported in part by COST Action CA16201 (PARTICLEFACE).
This work is supported by the Spanish Government  (Agencia Estatal de Investigaci\'on) and ERDF funds from European
Commission (Grants No. FPA2017-84445-P), Generalitat Valenciana (Grant No. PROMETEO/2017/053). 
JP acknowledges support from "la Caixa" Foundation (No. 100010434, LCF/BQ/IN17/11620037), and the European Union's H2020-MSCA 
Grant Agreement No. 713673; SRU from CONACyT and Universidad Aut\'onoma de Sinaloa; JJAV from Generalitat Valenciana (GRISOLIAP/2018/101); AERO from the Spanish Government (PRE2018-085925); and RJHP from the project A1-S-33202 (Ciencia B\'asica), Ciencia de Frontera 2021-2042 and Sistema
Nacional de Investigadores.


\bibliographystyle{JHEP}
\providecommand{\href}[2]{#2}\begingroup\raggedright\endgroup

\end{document}

%% file: draft_Andres.tex
Among the several advantages of the LTD formalism, we have recently explored the manifestly causal representations. This constitutes a big step towards an efficient numerical implementation of this approach, since spurious non-physical and non-causal singularities are completely avoided. In the following, we will introduce some useful notation and describe the basis of the formalism. In particular, we will explain how the LTD computation through the nested residue strategy leads to the causal representations of different topological families to all-loop orders. 

\subsection{Dual scattering amplitudes}
A generic $L$-loop scattering amplitude with $N$ external legs $\{p_j\}_N$, and $n$  sets of internal lines, each set being defined by the specific dependence on the loop momenta, is written as
\begin{equation} \label{AmpNL1}
    \mcal{A}_N^{(L)}(1,\dots,n) = \int_{\ell_1,\dots,\ell_L} 
    \mcal{A}_F^{(L)}(1,\dots,n)~,
\end{equation}
with 
\begin{equation} \label{AmpNL2}
    \mcal{A}_F^{(L)}(1,\dots,n) = \mcal{N}\big( \{ \ell_s\}_L, \{p_j\}_N \big) \times G_F(1,\dots,n) ~.
\end{equation}
This corresponds to an integral in the Minkoswki space of the $L$-loop momenta, $\{\ell_s\}_L$, involving the product of Feynman propagators, $G_F(q_i) = (q_i^2-m_i^2+\iz)^{-1}$, and numerators, $\mcal{N}(\{\ell_s\}_L,\{p_j\}_N)$, given by the Feynman rules of the theory considered. The $d$-dimensional integration measure in DREG reads
\beq 
\int_{\ell_s} = -\imath \, \mu^{4-d} \int {\rm d}^d\ell_s/(2\pi)^d \, ,
\eeq
and the usual Feynman propagator for a single particle is
\be
    G_F(q_{i_s}) = \frac{1}{q_{i_s,0}^2 - \LL(q_{i_s,0}^{(+)}\RR)^2} \, ,
\ee
where 
\be
    q_{i_s,0}^{(+)} = \sqrt{\mathbf{q}_{i_s}^2 + m_{i_s}^2 - \iz} ~, 
\ee
is the positive on-shell energy of the loop momentum $q_{i_s}$, written in terms of its spatial components $\mathbf{q}_{i_s}$, its mass $m_{i_s}$, and the Feynman's infinitesimal imaginary prescription $\iz$. Then, we introduce the shorthand notation 
\be \label{gfn}
    G_F(1,\dots,n) = \prod_{{i_s}\in1\cup\dots\cup n} \Big( G_F\big(q_{i_s}\big) \Big)^{a_{i_s}} ~,
\ee
to express the product of Feynman propagators of one set or the union of several sets. In the former formula, $s$ represents the set of all internal propagators with internal momenta $q_{i_s} = \ell_s+k_{i_s}$, 
that depend on the loop momentum, $\ell_s$, or a specific linear combination of loop momenta, and a combination of external momenta $k_{i_s}$, with $i_s\in s$ and $a_{i_s}$ arbitrary powers.
 
The LTD representation for scattering amplitudes is obtained by the iterative application of the Cauchy's residue theorem (CRT) integrating out one degree of freedom for each loop momentum, and closing the Cauchy contour from below the real axis, selecting the poles with negative imaginary part in the complex plane of the loop momentum. This results in the modification of the infinitesimal complex prescription of the Feynman propagators \cite{Catani:2008xa,Rodrigo:2008fp}, that needs to be considered carefully in order to preserve causality. Then, starting from \eq{AmpNL1}, we set on-shell the propagators that depend on the first loop momentum, $q_{i_1}$, and define 
\be
    \mcal{A}_D^{(L)}(1;2,\dots,n) \equiv 
    \sum_{i_1\in1}{\rm Res} \LL( \mcal{A}_F^{(L)}(1,\dots,n),\,{\rm Im}(q_{i_1,0}<0)\RR) \, ,
\ee
where taking the residue is equivalent to integrating out the energy component of the loop momenta. Now, we construct the nested residue iterating until the $r$-th set as,
\be \label{nesres}
    \mcal{A}_D^{(L)}(1,\dots, r \,;\, r+1, \dots, n) = 
    \sum_{i_s\in s} {\rm Res}\LL( \mcal{A}_D^{(L)}(1, \dots, r-1 \,;\, r,\dots,n),\, \text{Im}(q_{i_s,0}<0) \RR) ~.
\ee
All sets before the semicolon contain one propagator that has been set on shell and are linearly independent, while all the remaining propagators are kept off-shell. Thus, this representation is equivalent to open the loop amplitude to non-disjoint trees. Finally, the integration measure after integrating out the energy component is modified according to
\be
    \int_{\,\vec{\!\ell}_s} \equiv -\mu^{d-4} \int \frac{{\rm d}^{d-1}\ell_{s}}{(2\pi)^{d-1}} ~,
\ee
transforming the $d$-dimensional Minkowski space into a $(d-1)$-dimensional Euclidean one.

\subsection{Multi-loop topologies through the LTD}
\label{ssec:MultiLoopIntro}
\begin{figure}[t]
\centering
\includegraphics[scale=0.9]{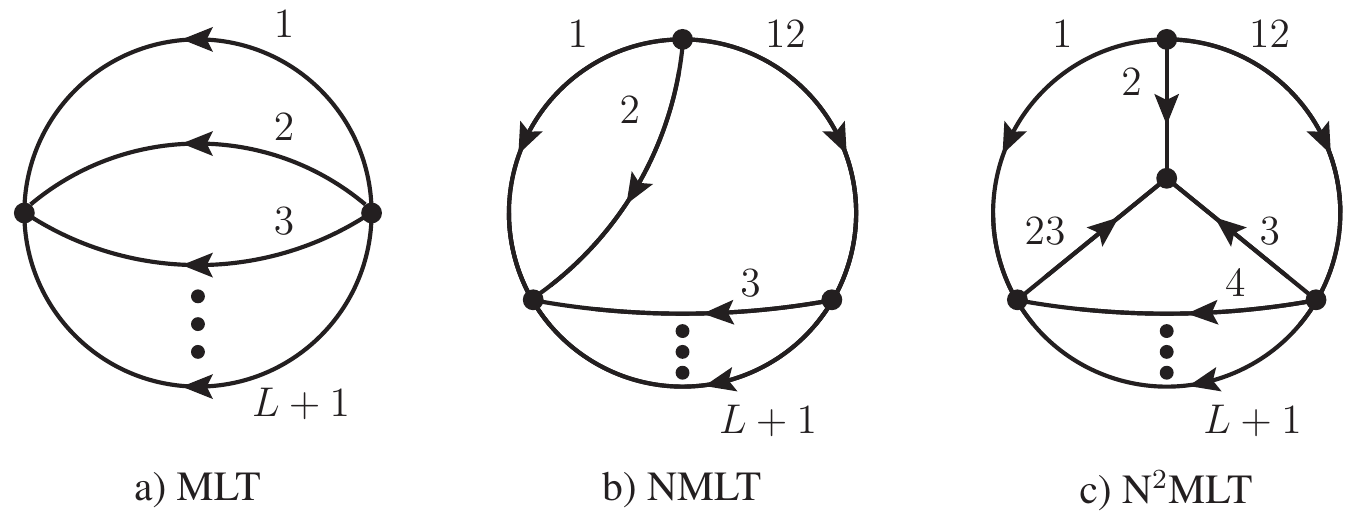}
\caption{Graphical representations of Maximal Loop Topology (MLT), Next-to-Maximal Loop Topology (NMLT) and Next-to-Next-to-Maximal Loop Topology (N$^2$MLT). External momenta are not shown.}
\label{fig:alltopos}
\end{figure}

By the properties obtained from the nested residues, \eq{nesres}, we are able to construct multi-loop amplitudes from the LTD representation which are compact and manifestly causal to all orders. The first topology under consideration is called Maximal Loop Topology (MLT), and is characterised by $L+1$ momentum sets where the momenta of the first $L$ sets depend on one single loop momentum
\beq 
q_{i_s}=\ell_s+k_{i_s} \, ,
\eeq
with $s\in\{1,\dots,L\}$. The momenta of the extra set, $L+1$, are given by a linear combination of all the loop momenta, namely $q_{i_{L+1}} = -\sum_{s=1}^{L}\ell_s+k_{i_{L+1}}$. Here, $k_{i_s}$ and $k_{i_{L+1}}$ denote linear combinations of external momenta.
The LTD representation of the MLT topology, displayed in Fig. \ref{fig:alltopos}a, is 
presented in a compact and symmetric form by evaluating the nested residue of \eq{nesres}, which leads to
\be \label{ampmlt}
    \mcal{A}_{\rm MLT}^{(L)}(1,\dots,L+1) = \int_{\vec{\ell}_1,\dots,\vec{\ell}_L}
    \sum_{i=1}^{L+1} \ \mcal{A}_D^{(L)}(1,\dots,i-1,\overline{i+1},\dots,\overline{L+1}\,;\, i)~,
\ee
where the bars within the integrand indicate a reversal of momentum flow, $q_{\bar{i}_s} = -q_{i_s}$, which is needed to preserve causality, and is equivalent to select the on-shell modes with negative energy of the original momentum flow. Each term of the sum in the integrand of \eq{ampmlt} contains one set $i$ with all its propagators off-shell, while the remaining $L$ sets contain a single on-shell propagator each; a necessary condition to open multi-loop amplitudes into disjoint trees. The dual representation of \eq{ampmlt} becomes singular when one or more off-shell propagators eventually become on-shell, generating a disjoint tree dual sub-amplitude.

\begin{figure}[t]
    \centering
    \includegraphics[scale=0.9]{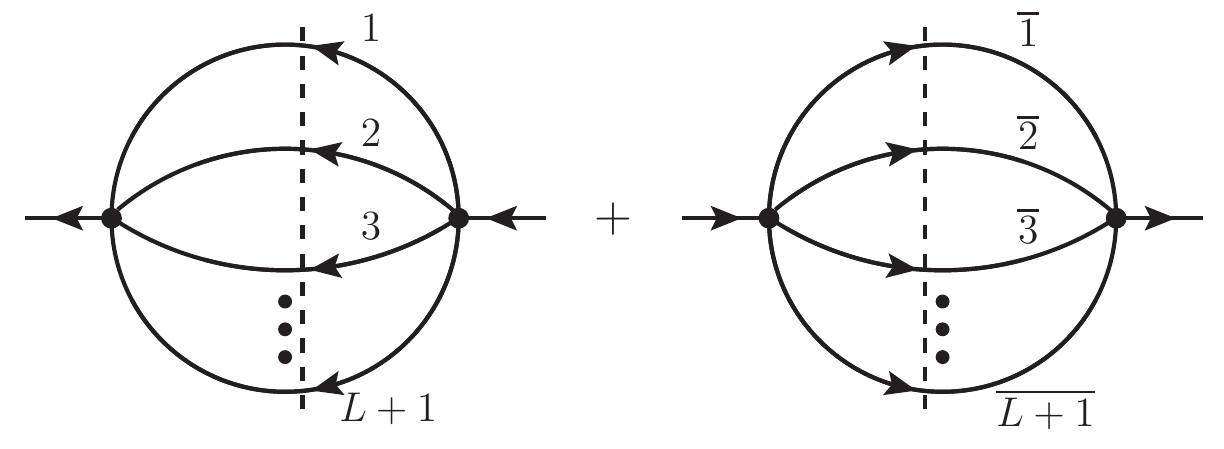}
    \caption{Representation of the causal structure of MLT topologies.}
    \label{fig:mlt}
\end{figure}

It is worth mentioning that the LTD representation exhibits an interesting structure when all the contributions are added together. For example, in the case of a MLT configuration with one propagator in each loop set and one incoming and outgoing momentum, we obtain
\begin{equation} \label{eq:fullmltp}
    {\cal A}_{\text{MLT}}^{\left(L\right)}\left(1,\dots,L+1\right) =
    -\int_{\vec{\ell}_1,\dots,\vec{\ell}_L}
    \frac{1}{x_{L+1}} \left(\frac{1}{\lambda_1^{-}}+\frac{1}{\lambda_1^{+}}\right) ~,
\end{equation}
with
\begin{align}
    x_{L+k}=2^{L+k}\prod_{i=1}^{L+k}q_{i,0}^{(+)} ~,
    \quad ~{\rm and}~ \quad
    \lambda_1^\pm = \sum_{i=1}^{L+1}q_{i,0}^{(+)} \pm k_{0,L+1} ~.
\end{align}
This expression is free of unphysical singularities, and written in terms of the so-called \emph{causal propagators}, $\lambda_{1}^{\pm}$. The causal propagators encode all the possible physical singularities that might occur. Both $\lambda_{1}^{\pm}$ in the integrand of \eq{eq:fullmltp} are associated to physical thresholds, as shown in Fig. \ref{fig:mlt}. Additionally, \eq{eq:fullmltp} is independent of the initial momentum configuration in the 
Feynman representation.

Going a step further in the topology complexity, we consider a topology containing one extra set of momenta that depends on the sum of two loop momenta, $q_{i_{12}} = -(\ell_1+\ell_2)+k_{i_{12}}$, denoted as 12. This configuration, as can be appreciated in Fig. \ref{fig:alltopos}b, is called Next-to-Maximal Loop Topology (NMLT), characterised by $L+2$ sets of propagators, with each set categorised by the dependence on a specific loop momentum or a linear combination of the $L$ independent loop momenta.
The LTD representation for the NMLT is given by the compact and factorised expression
\beqn \label{eq:nmlt}
  \mcal{A}_{\rm NMLT}^{(L)}(1,\dots,L+2) &=& \mcal{A}_{\rm MLT}^{(2)}(1,2,12) 
  ~\otimes~ \mcal{A}_{\rm MLT}^{(L-2)}(3,\dots,L+1)  \nn\\
  &+& \mcal{A}_{\rm MLT}^{(1)}(1,2) ~\otimes~ \mcal{A}^{(0)}(12)
  ~\otimes~ \mcal{A}_{\rm MLT}^{(L-1)}(\overline{3},\dots,\overline{L+1}) ~, 
\eeqn
involving convolutions of MLT subtopologies. The singular structure of the factorised subtopologies determine the causal thresholds and infrared singularities. For example, taking $L=3$, i.e. three loops, the convolutions within the NMLT configuration are interpreted as
\begin{align}
    \mathcal{A}_{\rm MLT}^{(2)}(1,2,12) \otimes \mathcal{A}_{\rm MLT}^{(1)}(3,4) &= 
    \int_{\vec{\ell}_1,\vec{\ell}_2,\vec{\ell}_3} 
    \Big( \mathcal{A}_D^{(3)}(\overline{2},\overline{12},\overline{4};1,3) 
        + \mathcal{A}_D^{(3)}(1,\overline{12},\overline{4};2,3)\nn\\ 
        &+ \mathcal{A}_D^{(3)}(1,2,\overline{4};12,3) + ( \overline{4}\leftrightarrow3 ) 
    \Big) ~,
\end{align}
and 
\begin{align}
    \mathcal{A}_{\rm MLT}^{(1)}(1,2) \otimes 
    \mathcal{A}^{(0)}(12) \otimes 
    \mathcal{A}_{\rm MLT}^{(2)}(\overline{3},\overline{4}) &=
    \int_{\vec{\ell}_1,\vec{\ell}_2,\vec{\ell}_3} 
    \Big( 
        \mathcal{A}_D^{(3)}(\overline{2},\overline{3},\overline{4};1,3) \nn\\
    &+ 
        \mathcal{A}_D^{(3)}(1,\overline{3},\overline{4};2,12) 
    \Big) ~,
\end{align}
where the sets after the semicolon are put off shell. 

Following the previous procedure, we also presented in Ref. \cite{Verdugo:2020kzh} the LTD representation for the 
Next-to-Next-to-Maximal Loop Topology (N$^2$MLT), characterised by $L+3$ sets of propagators and depicted in Fig. \ref{fig:alltopos}c. Again, factorised formulae for the dual representation are obtained, namely
\begin{align} \label{amp:n2mlt}
  \mathcal{A}_{\rm N^2MLT}^{(L)}(1, \dots , L+3, 12,23) &=
  \mathcal{A}_{\rm NMLT}^{(L)}(1,2,3,12,23) \otimes
  \mathcal{A}_{\rm MLT}^{(L-3)}(4,\dots,L+1) \nn\\ 
  &+ \mathcal{A}_{\rm MLT}^{(2)}(1\cup23,2,3\cup12) \otimes 
  \mathcal{A}_{\rm MLT}^{(L-2)}(\overline{4},\dots,\overline{L+1}) \, ,
\end{align}
showing a recursive construction similar to the one reported for the NMLT configurations. Expanding the first term on the rhs of \eq{amp:n2mlt} into its corresponding subtopologies, we have 
\begin{align}
    \mathcal{A}_{\rm NMLT}^{(L)}(1,2,3,12,23) 
&=  \mathcal{A}_{\rm MLT}^{(2)}(1,2,12) 
    \otimes \mathcal{A}_{\rm MLT}^{(3)}(3,23) \nn\\
&+  \int_{\vec{\ell}_1,\vec{\ell}_2,\vec{\ell}_3}
    \Big( \mathcal{A}_D^{(3)}(1,\overline{3},\overline{23};2,12) 
    + \mathcal{A}_D^{(3)}(\overline{12},3,23;1,2) \Big) \, ,
\end{align}
where the last two terms of the rhs in the integrand remain fixed by the condition that the sets (2,3,23) cannot generate a disjoint subtree. We can observe that the second term in \eq{amp:n2mlt}, contain a two-loop subtopology involving five sets of momenta grouped into three sets. Therefore, the propagators in the set 1 and 23 cannot be simultaneously off shell. 

Noteworthy, there are very compact explicit formulae for the NMLT and N$^2$MLT configurations which make use of the causal propagators. In Sec. \ref{sec:PaperWilliam}, we will enter into more details, and provide a nice conceptual interpretation of the manifestly causal dual representation in terms of entangled thresholds.

%% file: draft_JJ.tex
Generic scattering amplitudes are defined by integrals of rational functions. As already explained in Sec. \ref{sec:PaperPRL}, the multi-loop LTD framework is based on the CRT and, in this section, the formal foundations for an $L$-loop Feynman diagram and some of their immediate consequences are explained. As it was discussed in Ref. \cite{Aguilera-Verdugo:2020nrp}, iterated residues can be easily computed taking into account the quadratic structure of Feynman propagators. Let us start with the space $\mathbb{C}^{\left(\mathbb{R}^L\right)}$ of the functions with domain $\mathbb{R}^L$ and co-domain $\mathbb{C}$. If the natural inclusion of $\mathbb{R}$ into $\mathbb{C}$ is denoted by $i$, then the iterated residues are defined as the recursive application of the functor $-\mathrm{Res}\circ i:\mathbb{C}^{\left(\mathbb{R}^n\right)}\to\mathbb{C}^{\left(\mathbb{R}^{n-1}\right)}$, where $\mathrm{Res}$ represents the residue of the argument for all its negative imaginary part poles and the minus sign appears in agreement with CRT as the integration is performed clockwise.

\begin{figure}[H]
    \centering
    \includegraphics[width=0.35\textwidth]{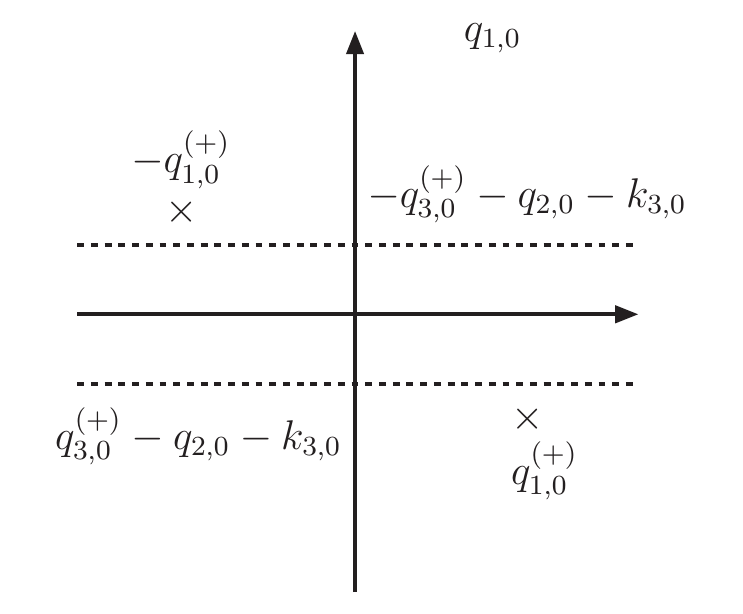}
    \caption{Pole structure of a rational function of two variables.}
    \label{fig:poles1}
\end{figure}

For the computation of multi-loop Feynman integrals and scattering amplitudes, internal momenta are given as linear combinations of external and loop momenta. For instance, for the scalar sunrise diagram, the associated integrand has the form \begin{equation}
    \mathcal{I}(p)=\frac{1}{\left(q_{1,0}^2-q_{1,0}^{(+)2}\right)\left(q_{2,0}^2-q_{2,0}^{(+)2}\right)\left((q_{1,0}+q_{2,0}+k_{3,0})^2-q_{3,0}^{(+)2}\right)}.
\end{equation}
In this case, the first iteration of the iterated residues can be developed with respect to the variable $q_{1,0}$. This means that we extend $q_{1,0}$ to the complex plane, and as $q_{2,0}$ is still a real variable, the poles are located in the complex plane as shown in Fig.~\ref{fig:poles1}, where the dashed line represents the poles $\pm q_{3,0}^{(+)}-q_{2,0}-k_{3,0}$. As they depend on $q_{2,0}$, they can be located at some point along each of the dashed lines.

\begin{figure}[H]
    \centering
    \includegraphics[width=0.8\textwidth]{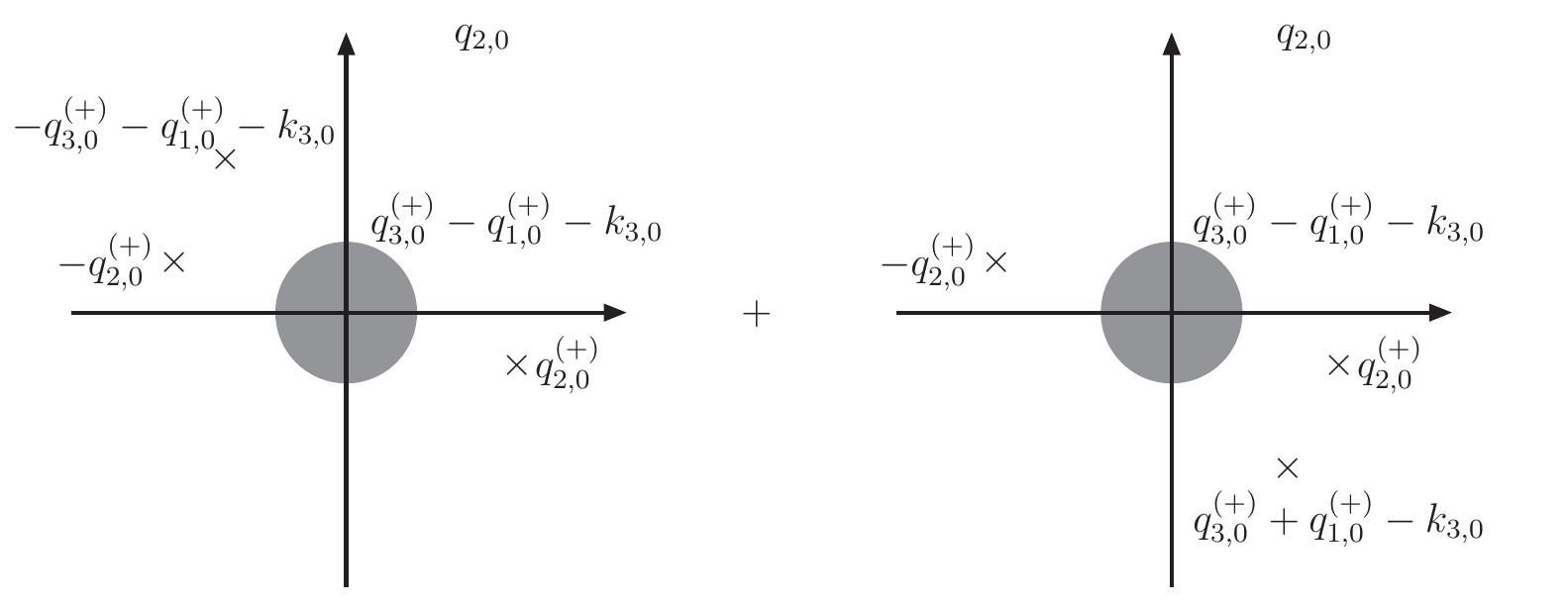}
    \caption{Pole structure of the first residue of a rational function of two variables.}
    \label{fig:poles2}
\end{figure}

Then, after the computation of the first iterated residues, we obtain \begin{equation}\label{eqn:1ressunrise}
    \begin{split}
        \mathrm{Res}[\mathcal{I}(p),\{q_{2,0},\mathrm{Im}(q_{2,0})<0\}]&=\frac{1}{2q_{1,0}^{(+)}\left(q_{2,0}^2-q_{2,0}^{(+)2}\right)\left(\left(q_{1,0}^{(+)}+q_{2,0}+k_{3,0}\right)^2-q_{3,0}^{(+)2}\right)}\\
        &+\frac{1}{2q_{3,0}^{(+)}\left(\left(q_{3,0}^{(+)}-q_{2,0}-k_{3,0}\right)^2-q_{1,0}^{(+)2}\right)\left(q_{2,0}^2-q_{2,0}^{(+)2}\right)} \, .
    \end{split}
\end{equation}
The function obtained after the first residue iteration is given as a sum of terms. The next step corresponds to the extension of the variable $q_{2,0}$ to the complex plane. There, the pole structure of each term of \Eq{eqn:1ressunrise} is quite similar. This can be represented pictorially as shown in the Fig.~\ref{fig:poles2}, where the pole structure of each term is presented as a complex plane term. The gray blob in both terms represents the dependence of the imaginary part of the pole $q_{3,0}^{(+)}-q_{2,0}^{(+)}-k_{3,0}$ on the three-momenta $\boldsymbol{q}_{2}$ and $\boldsymbol{q}_{3}$, having no definite imaginary part sign.

These poles having no definite imaginary part sign are, in general, called \emph{displaced poles}, and they happen to cancel their contributions to the iterated residues through the relation
\begin{equation}
    \begin{split}
        \mathrm{Res}&\left(\mathrm{Res}\left(F(q_{i,0},q_{j,0}),\left\{q_{i,0},q_{i,0}^{(+)}+a_i\right\}\right),\left\{q_{j,0},q_{k,0}^{(+)}-q_{i,0}^{(+)}+a_{ij}-a_i\right\}\right)\\
        =-&\mathrm{Res}\left(\mathrm{Res}\left(F(q_{i,0},q_{j,0}),\left\{q_{i,0},q_{k,0}^{(+)}-q_{j,0}+a_{ij}\right\}\right),\left\{q_{j,0},q_{k,0}^{(+)}-q_{i,0}^{(+)}+a_{ij}-a_i\right\}\right)
    \end{split}\label{eq:MasterCancellation}
\end{equation}
with
\beq
F(q_{i,0},q_{j,0}) =  \frac{P(q_{i,0},q_{j,0})}{\left((q_{i,0}-a_i)^2-q_{i,0}^{(+)2}\right)^{\gamma_i}\left((q_{i,0}+q_{j,0}-a_{ij})^2-q_{k,0}^{(+)2}\right)^{\gamma_k}} \, ,
\label{eq:FuncionEspecial}
\eeq for $a_i$ and $a_{ij}$ linear combinations of energy components of external momenta. Due to this cancellation, displaced poles can be ignored in the computation, leading to the concept of \emph{nested residues}. Furthermore, the final result of the nested residues is independent of the order of the iteration, although the expressions are not identical term by term. A rigorous proof of \eq{eq:MasterCancellation} is shown in Ref. \cite{Aguilera-Verdugo:2020nrp}.

As a consequence of the nested residue strategy, we can infer a general formula for the causal structure of the MLT diagram (see Fig.~\ref{fig:alltopos}a).
This is directly reached applying partial fractions to each term in the nested residue, leading to \begin{equation}\label{eqn:mltcausal}
    \begin{split}
        G_F(1,\hdots,L+1)\to-\frac{1}{\prod\limits_{i=1}^{L+1}2q_{i,0}^{(+)}}\left(\frac{1}{\sum\limits_{i=1}^{L+1}q_{i,0}^{(+)}+p_0}+\frac{1}{\sum\limits_{i=1}^{L+1}q_{i,0}^{(+)}-p_0}\right).
    \end{split}
\end{equation}
Additionally, this result can be used for the insertion of an MLT diagram within a more general topology. In this manner, it is possible to consider any MLT insertion as a single propagator whose on-shell energy is the sum of the on-shell energies of all the internal propagators of the MLT.

\begin{figure}[H]
    \centering
    \includegraphics[width=0.65\textwidth]{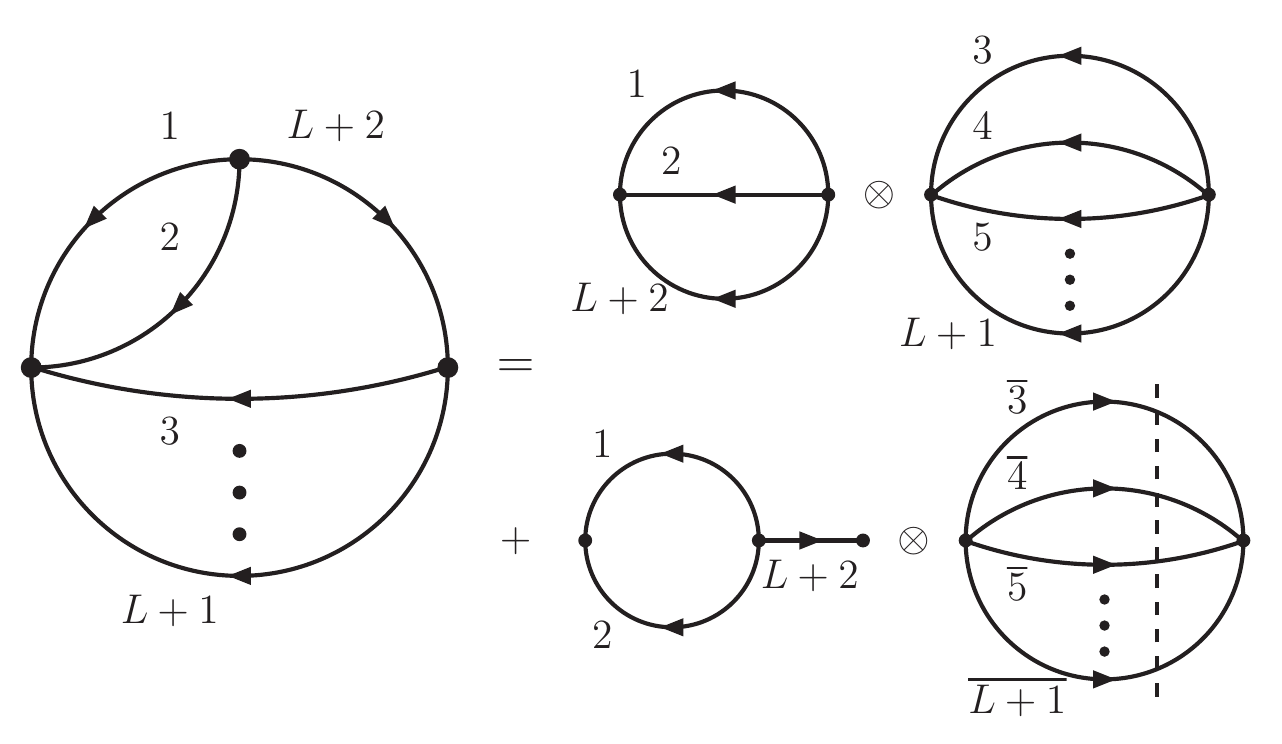}
    \caption{Dual decomposition of NMLT$(L)$ in terms of loop configurations with lower topological complexity.}
    \label{fig:NMLTdual}
\end{figure}

\begin{figure}[H]
    \centering
    \includegraphics[width=0.65\textwidth]{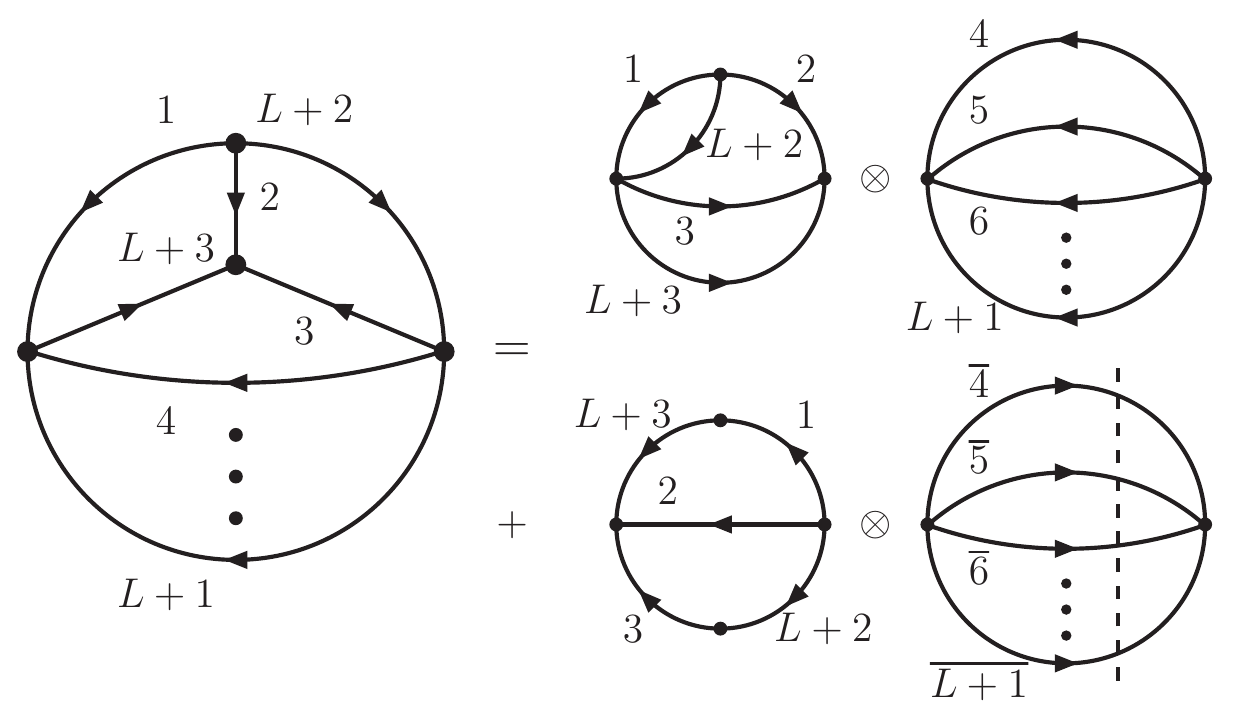}
    \caption{Dual expansion of a N$^2$MLT$(L)$ diagram.}
    \label{fig:NNMLTdual}
\end{figure}

Furthermore, with the computation of the nested residues, every family of a given topological complexity $k$, N$^{k-1}$MLT, can be re-expressed as a linear combination of convolutions of lower topological complexity families, as it is for the NMLT and N$^2$MLT, both defined by their amplitudes, \begin{equation}
    \begin{split}
        \mathcal{A}_{\mathrm{NMLT}}^{(L)}(1,\hdots,L+2)&\equiv\int\limits_{\ell_1,\hdots,\ell_L}\mathcal{N}(\{\ell_i\}_{L},\{p_j\}_N)\times G_F(1,\hdots,L+1,12),\\
        \mathcal{A}_{\mathrm{N}^2\mathrm{MLT}}^{(L)}(1,\hdots,L+3)&\equiv\int\limits_{\ell_1,\hdots,\ell_L}\mathcal{N}(\{\ell_i\}_{L},\{p_j\}_N)\times G_F(1,\hdots,L+1,12,23).
    \end{split}
\end{equation}
This decomposition is shown in Figs.~\ref{fig:NMLTdual} and~\ref{fig:NNMLTdual} for each of these topological families.

It is worth mentioning that the convolution symbol does not represent a pure factorisation, as it implies the use of the on-shell conditions to express all the off-shell momenta. The NMLT($L$) diagram decomposes into two terms with two MLT diagrams. One of these terms contains a MLT($L-2$) in convolution with a MLT(2), while the other term contains a convolution of a MLT($L-2$) diagram with all its internal lines on-shell with a MLT(1) diagram and one off-shell momentum. In the case of the N$^2$MLT diagram, it is also decomposed into two terms. One of these terms is a convolution of an NMLT(3) diagram with a MLT($L-3$) diagram, and the other term is a convolution of one fully on-shell MLT($L-3$) diagram with a MLT(3) diagram with two external momenta inserted in the first and third internal set. These external momenta play a fundamental role in splitting the internal sets into two propagators each. In this manner, the following relations are fulfilled
\begin{align}\label{eqn:topfam}
        \mathrm{NMLT}(L)&=\mathrm{MLT}(2)\otimes \mathrm{MLT}(L-2)+G_F(L+2)\otimes\mathrm{MLT}(1)\otimes \mathrm{MLT}(L-2) \, ,\nn\\
        \mathrm{N}^2\mathrm{MLT}(L)&=\mathrm{NMLT}(3)\otimes\mathrm{MLT}(L-3)+\mathrm{MLT}(2)\otimes\mathrm{MLT}(L-3) \, ,
\end{align}
which justifies the factorisation relations presented in Sec. \ref{sec:PaperPRL}.

The topological complexity of the subdiagrams in the convolutions is an additive quantity whose sum coincides with the topological complexity of the original diagram. Recalling that the family of Feynman diagrams with topological complexity $k$ with $L$ loops is N$^{k-1}$MLT($L$), for the NMLT(L) has $k_{\rm NMLT}=2$ and the term in the decomposition with no fully on-shell diagrams is given as the convolution of MLT diagrams (with $k_{\rm MLT}=1$), thus $2k_{\rm MLT}=k_{\rm NMLT}$. For the N$^2$MLT($L$) it is given $k_{\rm N^2MLT}=3$, while the term in its decomposition with no fully on-shell diagrams is the convolution of an NMLT($L$) and a MLT($L$) diagram, hence, $k_{\rm NMLT}+k_{\rm MLT}=k_{\rm N^2MLT}$.

These results suggest a factorisation formula for each topological complexity family. Indeed, these formulae can be reached through the nested residues and reveal a relation between one topological family with topologies with lower complexity. Within such an approach, every topological family can be expanded by factorising the MLT diagrams; thus, for any topological family, its causal structure can be expressed following \eq{eqn:mltcausal}. We profited from these structures to derive all-order formulae for N$^3$MLT and N$^4$MLT families, as explained in Sec. \ref{sec:PaperSelo}.

%% file: draft_Judith.tex
The LTD formalism is straightforwardly applicable to obtain several simplifications of the integrand-level representations of loop amplitudes, such as asymptotic expansions. The interest in asymptotic expansions arises from their potential to facilitate analytic results in specific kinematic configurations, particularly when full analytic calculations are not (yet) possible. An asymptotic expanded result can be of great interest since it showcases the relevant behaviour of the full amplitude in the needed kinematic limit. There are many observables where an analytic result is not necessary for every set of kinematics and where specific limits are the window to test potential discrepancies between experimental measurements and SM predictions, thus identifying new physics contributions. Furthermore, in the context of the local cancellation of IR singularities, expanded integrands could be very convenient to reduce computation time. While maintaining the correct analytic structure in the divergent limit and thus allowing for the combination with the real-emission contributions, the less complicated form of the expanded virtual contributions is expected to evaluate faster during the point-by-point process of numerical integration.

Since after the application of LTD an amplitude is reduced to an integral over Euclidean three-momenta, the size of the appearing scalar products can be unambiguously be compared to external scales. This provides a good starting point for the development of a well-defined formalism for asymptotic expansions of the integrand. Specific asymptotic expansions in the context of LTD have been presented for the first time for the process $H\to\gamma\gamma$ at one loop \cite{Driencourt-Mangin:2017gop}. Recently advances in the generalization of the formalism have been published in Refs. \cite{Plenter:2019jyj,Plenter:2020lop}.

Due to the fact that scattering amplitudes are determined by their analytic properties general considerations for integrand-level expansions should start with the propagators as the origin of divergences. The numerator, while playing a role in the appearance of \uv divergences, is not relevant for the discussion of asymptotic expansions since within LTD the singular \uv~ behaviour is neutralised through local renormalisation before integration.

We can reparametrise the dual propagators in the following form that 
is more suitable for asymptotic expansions
\begin{equation}
\td{q_i} \, G_D \left( q_i ; q_j\right) = \frac{\td{q_i}}{2q_i\cdot k_{ji} + \Gamma_{ij}+\Delta_{ij} - \ii \eta\cdot k_{ji}}~,
\label{eq:dualpropagator}
\end{equation}
where $\Gamma_{ij}+\Delta_{ij} = k_{ji}^2 + m_i^2 - m_j^2$. If $\Gamma_{ij}+\Delta_{ij}$ vanishes, the dual propagator is not expanded. Otherwise the starting point for the asymptotic expansion is to demand that the condition
\begin{equation}
|\Delta_{ij}| \ll |2q_i\cdot k_{ji} + \Gamma_{ij}| 
\label{eq:condition_convergence}
\end{equation}
be fulfilled for the whole range of the loop integration space except for potentially small regions around physical divergences. Note that since dual propagators only appear in integrands where one loop momentum has been set on shell, the condition has to be fulfilled in the Euclidean space of the loop three-momentum. The dual propagator can then be expanded as
\begin{equation}
G_D \left( q_i ; q_j\right)= \sum_{n=0}^\infty \frac{\left(-\Delta_{ij} \right)^n}{\left(2q_i\cdot k_{ji} + \Gamma_{ij} - i0 \eta\cdot k_{ji}\right)^{n+1}}~. 
\label{eq:expandedpropagator}
\end{equation}
Further simplifications arise whenever ${\bf k}_{ji}=0$. In that case, with the change of variables $|\vec{q_i}| = m_i/2 \, (x_i - x_i^{-1})$, the denominator of the expanded dual propagator takes the easily integrable form
\begin{equation}
2q_i\cdot k_{ji} + \Gamma_{ij} - \ii \,\eta\cdot k_{ji} = Q_i^2 \left( x_i + r_{ij}\right)\left( x_i^{-1} + r_{ij} \right)~. 
\label{eq:expandedpropagator_denominator}
\end{equation}
The intention to rewrite the denominator in this shape determines the parameters $\Gamma_{ij}$ and $r_{ij}$ appearing in the expansion to be restricted by the conditions 
\begin{align}
& \Gamma_{ij} - \ii \,\eta\cdot k_{ji} = Q_i^2 \left( 1+ r_{ij}^2 \right)~, \label{eq:condition_coefficients} 
\\ & r_{ij} = \frac{m_i \, k_{ji,0}}{Q_i^2} -\frac{\ii \, \eta\cdot k_{ji}}{Q_i^2}~,
\end{align}
assuming $|r_{ij}|\le 1$.
In the types of limits where one hard scale $Q$ is available, $Q_i^2 = \pm Q^2$ can be identified. The sign is determined by the sign of the hard scale in the expression $k_{ji}^2+m_i^2-m_j^2$. This type of expansion facilitates the analytical integration based on integrals of the form 
\begin{equation}
\int_1^\infty \frac{\mathrm{d} x_i}{x_i (x_i + r_{ij})(x_i^{-1} +r_{ij})} = \frac{\log(r_{ij})}{r_{ij}^2-1}, \qquad ~ |r_{ij}|<1~. 
\label{eq:exampleintegral}
\end{equation}

In addition to the relations in \eq{eq:condition_coefficients} further conditions have to be respected by the expansion parameters to ensure that the expansion converges both at integrand- and at integral-level. Specifically, it is fundamental that the analytic behaviour of the dual propagator is not fundamentally changed, i.e. that for a propagator with a singularity the expansion also has to display that singularity, while the expansion of a non-singular propagator is to be finite throughout the whole integration domain. The infinitesimal imaginary prescription of $r_{ij}$ given in \eq{eq:condition_coefficients} accounts properly for the complex prescription of the original dual propagator and therefore its causal thresholds. A different approach has to be taken for threshold limits where a hard scale is not identifiable. In this case the pole position of the non-expanded propagator can be expanded to identify the correct $r_{ij}$ for the asymptotic expansion.

The formalism of expanding the dual propagator has been developed through its application on the locally renormalised scalar two-point function
\begin{equation}
 \mathcal{A}^{(1,R)} =  \int_\ell  \Big[ G_F(q_1; M) \, G_F(q_2; m) -  \left(G_F(\ell; \mu_\uv)\right)^2 \Big]~.
\end{equation}
Applying \Eq{eq:expandedpropagator} on the appearing dual propagators $G_D(q_1;\ell)$ and $G_D(\ell;q_1)$ leads to a simplified expression that can be integrated without needing to specify what type of limit is considered, giving the very general result
\begin{equation}
 \mathcal{A}^{(1,R)} =  \frac{1}{16\pi^2 } \sum_{i,j} \Bigg[ 2 + c_{0,i} \ln{\frac{\mu_\uv}{m_i}}  + \sum_{n=0}^\infty \left( c^{(n)}_{1,i} + c^{(n)}_{2,i}  \ln{r_{ij}} \right)  \Bigg]~. \label{eq:asymptotic_general}
\end{equation}
The coefficients $c$ are simple functions of the appearing scales and are given in Ref. \cite{Plenter:2020lop} just like the parameters needed for different limits. The expansion converges well both at integrand- and at integral-level in the limits of one large mass, a large external momentum as well as when approaching the physical threshold both from below and from above. Comparison with the established method of \textit{expansion by regions} \cite{Beneke:1997zp,Pak:2010pt,Jantzen:2011nz,Mishima:2018olh,Semenova:2018cwy} has shown faster convergence as well as emphasised the advantage of decreasing degree in UV divergence with each order in the expansion. Renormalisation within our method thus only involves the lowest orders of the integrand-level expansion. Higher terms are optional for increasing precision and can be added straight-forwardly without the need to ensure further UV cancellation.

\begin{figure}
	\centering
	\includegraphics[width=0.75\textwidth]{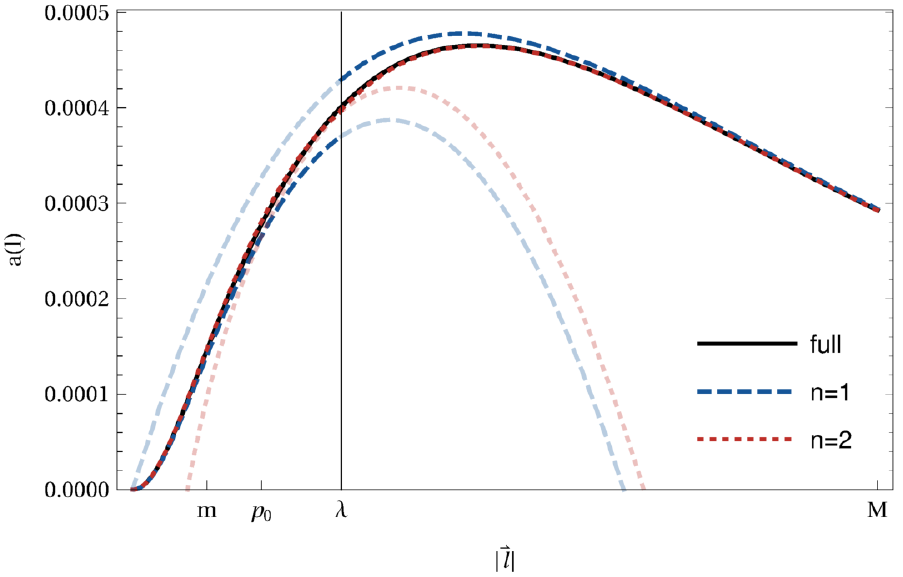}
	\caption{The convergence at integrand-level of the expansion given in \eq{eq:Taylor_expansion} for the values $M=10\, m$, $p^2=3\, m^2$, and $\mu_\uv=M$. More details about this plot can be found in the original paper \cite{Plenter:2020lop}.
	\label{fig:Taylor_integrand_convergence}}
\end{figure}

Using the same approach we have found integrand-level expansions for the scalar three-point function
\begin{equation}
{\cal A}_3^{(1)} = \int_\ell G_F(q_1,q_2,q_3; M)~,
\end{equation}
both for the limit of a large mass
\begin{equation}
{\cal A}_3^{(1)} ( s_{12} \ll M^2 )  
= \quad -\frac{1}{16\pi^2 } \frac{1}{2M^2} \left( 1+\frac{r}{12} + \frac{r^2}{90} \right) + {\cal O} (r^3)~ ,
\end{equation}
as well as for a small mass
\begin{align}
{\cal A}_3^{(1)} (s_{12} \gg M^2 ) =  &\, \int_\ell \frac{\td{\ell; M} s_{12}}{(2\ell\cdot p_{12})(2\ell\cdot p_1)}  \sum_{n=0}^\infty \Bigg\{ 
\frac{M^{2n}}{\left( -2\ell\cdot p_{12} +\Gamma\right)^{n+1}}
+ \frac{M^{2n}}{\left( 2\ell\cdot p_{12} +\Gamma\right)^{n+1}}\Bigg\}~. \label{eq:trianglesmallreg}  \nn \\
\end{align}
The Euclidean structure of the dual integrand, $a(\boldsymbol{\ell})$, can be exploited into an even more direct way by applying Taylor expansions. Here it is important to use different assumptions on the size of the loop three-momentum in separate parts of the integration range. For the scalar two-point function in the large-mass limit this amounts to calculating
\begin{equation}
\mathcal{A} = \int_0^\lambda\mathrm{d} |\boldsymbol{\ell}|~ \mathcal{T} a (M,\infty) + \int_\lambda^\infty\mathrm{d} |\boldsymbol{\ell}|~ \mathcal{T} a  (\{ \boldsymbol{\ell},M\}, \infty)~. \label{eq:Taylor_expansion}
\end{equation}
The expanded integrand converges well in the full range of the loop momentum as can be seen in Fig. \ref{fig:Taylor_integrand_convergence}. Details on this method and on how to fix the matching scale lambda that separates the appearing so-called dual regions can be found in \cite{Plenter:2020lop}.

With the calculations above being performed in the traditional LTD representation we will finally show an example of using the manifestly causal representation to expand a multi-loop integral. The MLT structure offers an ideal starting point, being void of unphysical singularities. For the large mass limit we find
\begin{equation}
{\cal A}^{(L)}_{\rm MLT}(p^2\ll m_s^2) = - 2 \sum_{n=0}^\infty  (p^2)^{n} \int_{\vec \ell_1, \ldots ,\vec \ell_L}  
\frac{\left( \lambda^0_{L+1} \right)^{-1-2n} }{x_{L+1}}~, 
\label{eq:asymptMLT}
\end{equation}
where $\lambda^0_{L+1} = \sum_{s=1}^{L+1} \qon{s}$ and $x_{L+1} = \prod_{s=1}^{L+1} 2q_{s,0}^{(+)}$. 
Notice that due to the lack of dependence on $p_0$ both in $x_{L+1}$
nor in  $\lambda^0_{L+1}$ the asymptotic integrals on the right-hand side of \eq{eq:asymptMLT} are a function only of the internal masses, $m_s$, to all loop orders. In this way, we are optimistic about future applications of the manifestly causal LTD representation from Refs. \cite{Verdugo:2020kzh,Aguilera-Verdugo:2020kzc,Ramirez-Uribe:2020hes,Aguilera-Verdugo:2020nrp} to speed up the calculation of efficient and smooth asymptotic expansions.

%% file: draft_GW.tex
As explained in Secs. \ref{sec:PaperPRL} and \ref{sec:PaperJesus}, the application of the nested residue strategy leads to manifestly causal integrand-level representations of any multi-loop multi-leg Feynman amplitude. The main advantage of such a representation is the absence of non-physical singularities, because only terms compatible with causality remain in the final result. 

Starting from the compact LTD representations of the NMLT and N$^2$MLT multi-loop topologies
presented in Ref.~\cite{Verdugo:2020kzh} in terms of nested residues, we reconstruct their full analytic expression in term of causal propagators only. This operation is performed by using numerical evaluation over finite fields~\cite{vonManteuffel:2014ixa,Peraro:2016wsq},
in which we use the \verb"C++" implementation of the \FiniteFlow~\cite{Peraro:2019svx} algorithm together with its \Mathematica ~interface. In the following Section, we show the explicit causal formulae for NMLT and N$^2$MLT families for an arbitrary number of loops, and discuss their interpretation in terms of entangled causal thresholds. Numerical examples for 4-loop diagrams are presented, in order to provide a comparison with available results and test the efficiency of our approach.

\subsection{Next-to-Maximal Loop Topology (NMLT)}
\label{ssec:NMLT}
At three-loops, the MLT family is not enough to describe the whole set of possible topologies. Thus we need to consider the NMLT and N$^2$MLT topologies, whose general dual representations were explained in Sec. \ref{sec:PaperPRL}. To simplify the presentation, we start by considering NMLT configurations with one single propagator in each set and no external momenta. We need to add an additional internal line, whose momentum is given by 
\begin{align}
q_{L+2}=-\ell_{1}-\ell_{2} \,.\label{eq:setnmlt}
\end{align}
A pictorial representation of NMLT is provided in Fig~\ref{fig:alltopos}\textcolor{blue}{b}. After computing the nested residues and adding all the contributions together, we get
\begin{align}
{\cal A}_{\text{NMLT}}^{\left(L\right)}\left(1,2,\hdots,L+2\right)=&\int_{\vec{\ell}_{1},\hdots,\vec{\ell}_{L}}\frac{2}{x_{L+2}}\left(\frac{1}{\lambda_{1}\lambda_{2}}+\frac{1}{\lambda_{2}\lambda_{3}}+\frac{1}{\lambda_{3}\lambda_{1}}\right)\,,
\label{eq:fullnmlt}
\end{align}
where the causal propagators are given by 
\begin{align}
\lambda_{1}=\sum_{i=1}^{L+1}q_{i,0}^{\left(+\right)}\, & & \lambda_{2}=q_{1,0}^{\left(+\right)}+q_{2,0}^{\left(+\right)}+q_{L+2,0}^{\left(+\right)}\,, 
& & \lambda_{3}=\sum_{i=3}^{L+2}q_{i,0}^{\left(+\right)}\,.
\label{eq:l1tol3}
\end{align}
This expression was reconstructed by using numerical evaluations over finite fields, although partial fractioning leads to the same result within a similar computing time. In the following, we shall note that simplifications are not straightforward when dealing with more complicated topologies.

\begin{figure}[t]
\centering
\includegraphics[scale=0.9]{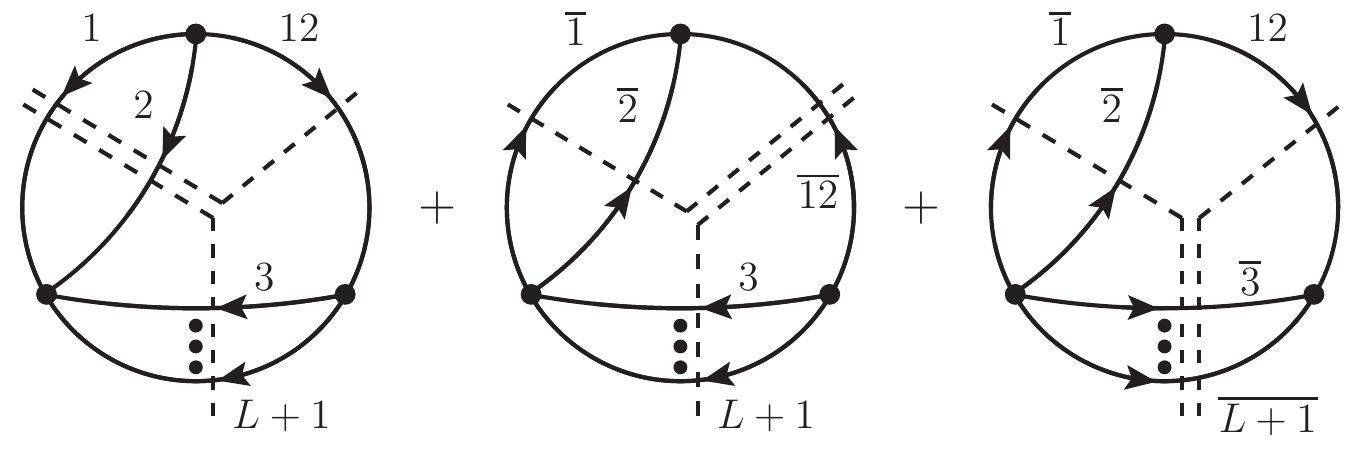}
\caption{Entangled causal thresholds of the NMLT topology. Products of two causal propagators are involved.}
\label{fig:nmlt}
\end{figure}

An interesting consequence of the compact form of \Eq{eq:fullnmlt} is that it allows a re-interpretation in in terms of entangled causal thresholds. Each $\lambda_i$ is associated to a causal threshold singularity, which might take place when the momenta flows are oriented in the same direction. Thus, the product of causal denominators represents a configuration in which two (or more) sets of propagators can simultaneously go on shell. The factorisation of NMLT (and more complicated topologies) as products of MLT configurations is the reason behind this behaviour~\cite{Verdugo:2020kzh}. A graphical interpretation of the entangled causal structure of NMLT vacuum diagrams is provided in Fig.~\ref{fig:nmlt}, with the dashed lines indicating the propagators that can be set simultaneously on shell.

\subsection{Next-to-Next-to-Maximal Topology (N$^{2}$MLT)}
\label{ssec:N2MLTvacuum}
As mentioned before, reaching a full description of N$^k$MLT with $k\leq2$ is enough to obtain the causal representation of up to three-loop scattering amplitudes. The so-called N$^2$MLT can be built recursively from the NMLT by adding an additional line with momentum
\begin{align}
q_{L+3}=-\ell_{2}-\ell_{3} \,.\label{eq:setn2mlt}
\end{align}
The minimal example of such topology is the Mercedes-Benz diagram ($L=3$) shown in Fig.~\ref{fig:alltopos}\textcolor{blue}{c}. For the sake of simplicity, we restrict here to the case without external momenta. Using the LTD representation in Ref.~\cite{Verdugo:2020kzh}, we can add together all the contributions and obtain
\begin{align}
{\cal A}_{\text{N}^2\text{MLT}}^{\left(L\right)}\left(1,2,\hdots,L+3\right) &= \int_{\vec{\ell}_{1},\hdots,\vec{\ell}_{L}}
\frac{1}{x_{L+3}}\frac{{\cal N}(\{q_{i,0}^{(+)}\})}{\prod_{i=1}^{7}\lambda_{i}}\,,
\end{align}
with $\lambda_1$ through $\lambda_3$ defined in \Eq{eq:l1tol3}, 
\begin{align}
 & \lambda_{4}=q_{2,0}^{\left(+\right)}+q_{3,0}^{\left(+\right)}+q_{L+3,0}^{\left(+\right)}\,, &  & \lambda_{6}=q_{1,0}^{\left(+\right)}+q_{3,0}^{\left(+\right)}+q_{L+2,0}^{\left(+\right)}+q_{L+3,0}^{\left(+\right)}\,,\nonumber \\
 & \lambda_{5}=q_{1,0}^{\left(+\right)}+q_{L+3,0}^{\left(+\right)}+\sum_{i=4}^{L+1}q_{i,0}^{\left(+\right)}\,, &  & \lambda_{7}=q_{2,0}^{\left(+\right)}+\sum_{i=4}^{L+3}q_{i,0}^{\left(+\right)}\,,
\end{align}
and ${\cal N}(\{ q_{i,0}^{(+)}\})$ a degree-four polynomial in $q_{i,0}^{(+)}$. Compared to the NMLT case, the complexity of the polynomial in the denominator makes it highly non-trivial to unveil a formula similar to \Eq{eq:fullnmlt}. Thus, we rely on the analytic reconstruction over finite fields to obtain 
\begin{align}
{\cal A}_{\text{N}^{2}\text{MLT}}^{\left(L\right)}\left(1,2,\hdots,L+3\right)=&-\int_{\vec{\ell}_{1},\cdots,\vec{\ell}_{L}}\frac{2}{x_{L+3}}\Bigg[\frac{1}{\lambda_{1}}\left(\frac{1}{\lambda_{2}}+\frac{1}{\lambda_{3}}\right)\left(\frac{1}{\lambda_{4}}+\frac{1}{\lambda_{5}}\right)\nonumber\\
&+\frac{1}{\lambda_{6}}\left(\frac{1}{\lambda_{2}}+\frac{1}{\lambda_{4}}\right)\left(\frac{1}{\lambda_{3}}+\frac{1}{\lambda_{5}}\right)+\frac{1}{\lambda_{7}}\left(\frac{1}{\lambda_{2}}+\frac{1}{\lambda_{5}}\right)\left(\frac{1}{\lambda_{3}}+\frac{1}{\lambda_{4}}\right)\Bigg]\,.
\label{eq:fulln2mlt}
\end{align}
It is worth appreciating that the package \textsc{Lotty} was used to efficiently reach these results \cite{TorresBobadilla:2021dkq}. As in the case of NMLT topologies, it is possible to interpret this result by using entangled thresholds. This time, there are products of three causal propagators. Notice that not all the combinations of causal propagators are allowed. This is because causal propagators exhibit some compatibility issues, which can be explained by digging into graph theory. More details about this issue and the connection with Cutkosky's rules can be found in Ref. \cite{Sborlini:2021owe}.

\begin{figure}[t]
\centering
\includegraphics[scale=0.9]{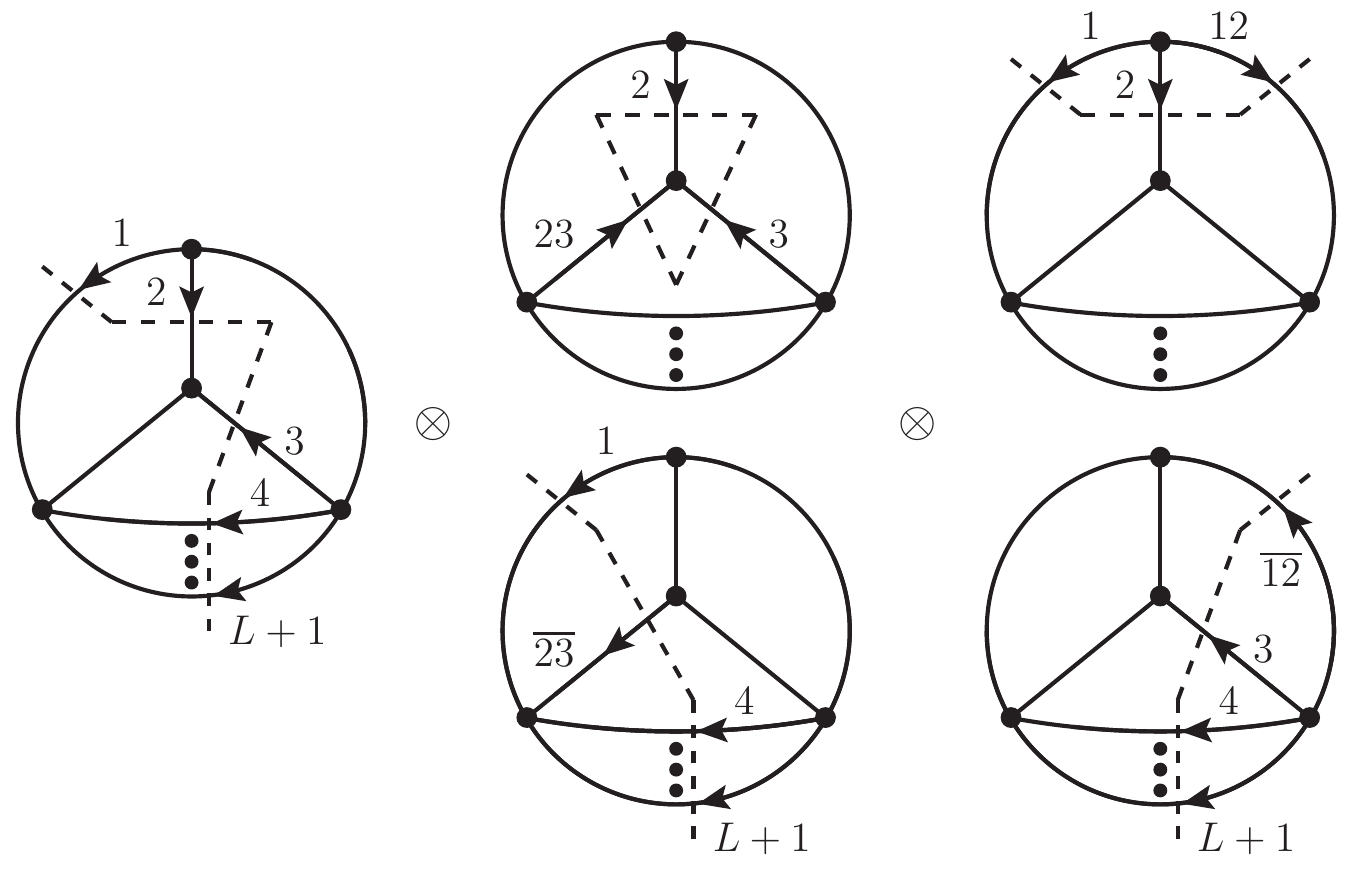}
\caption{Entangled causal thresholds of the N$^2$MLT topology. Products of three causal propagators are involved.}
\label{fig:n2mlt}
\end{figure}

\subsection{Adding external momenta and higher-powers}
\label{ssec:BeyondSimple}
For realistic multi-loop scattering amplitudes, we need to take into account external legs. The analytic reconstruction algorithm used to obtain compact formulae for vacuum diagrams can be also applied to topologies with external momenta. We want to highlight that the insertion of external momenta does not affect
the causal physical behaviour of these integrals: the difference w.r.t. the vacuum case is that the entangled configurations are duplicated, according to the direction of the energy flow for external particles. For instance, a generic N$^2$MLT with external legs inserted in the vertices is given by
\begin{align}
{\cal A}_{\text{N}^{2}\text{MLT}}^{\left(L\right)}\big(1,2 & ,\hdots,\left(L+1\right)_{-p_{13}},\left(L+2\right)_{p_{2}},\left(L+3\right)_{-p_{3}}\big)=-\int_{\vec{\ell}_{1},\cdots,\vec{\ell}_{L}}\frac{1}{x_{L+3}}\nonumber \\
\times & \Bigg[\frac{1}{\lambda_{1}^{+}}\left(\frac{1}{\lambda_{2}^{-}}+\frac{1}{\lambda_{3}^{-}}\right)\left(\frac{1}{\lambda_{4}^{+}}+\frac{1}{\lambda_{5}^{+}}\right)+\frac{1}{\lambda_{6}^{+}}\left(\frac{1}{\lambda_{3}^{-}}+\frac{1}{\lambda_{5}^{-}}\right)\left(\frac{1}{\lambda_{2}^{+}}+\frac{1}{\lambda_{4}^{+}}\right)\nonumber \\
 & +\frac{1}{\lambda_{7}^{+}}\left(\frac{1}{\lambda_{3}^{-}}+\frac{1}{\lambda_{4}^{-}}\right)\left(\frac{1}{\lambda_{2}^{+}}+\frac{1}{\lambda_{5}^{+}}\right)+\left(\lambda_{i}^{+}\leftrightarrow\lambda_{i}^{-}\right)\Bigg]\,.
\end{align}
More details about the algorithms used to perform this computations can be found in Refs.~\cite{Aguilera-Verdugo:2020kzc,TorresBobadilla:2021dkq}.

Besides that, it was shown in Refs. \cite{Hernandez-Pinto:2015ysa,Sborlini:2015uia,Sborlini:2016gbr,Sborlini:2016hat} that the presence of self-energy insertions or generic scattering amplitudes, as well as some local UV counter-terms, might require to consider propagators with higher-powers. As explained in Refs. \cite{Aguilera-Verdugo:2020kzc,Aguilera-Verdugo:2020nrp}, the causal structure of these amplitudes can be obtained by applying a differential operator. Explicitly, we can raise the power of the propagators by taking derivatives w.r.t. $q_{i,0}^{\left(+\right)}$. For instance, 
\begin{align}
\left( G_{F}\left(q_{i}\right) \right)^{\alpha_i}=\frac{1}{\left(\alpha_i-1\right)!}\frac{\partial^{\alpha_i-1}}{\partial\left((q_{i,0}^{\left(+\right)})^{2}\right)^{\alpha_i-1}}
\, G_{F} (q_i)\,,
\end{align}
which suggests the definition of the operator
\begin{align}
\frac{\partial}{\partial(q_{i,0}^{\left(+\right)})^{2}}\,\bullet & =\frac{1}{2q_{i,0}^{\left(+\right)}}\frac{\partial}{\partial(q_{i,0}^{\left(+\right)})}\,\bullet\,.
\end{align}
The iterated application of this operator to the causal representation of scattering amplitudes with single powers of the denominators leads to causal representation of the corresponding multi-power amplitude, as carefully explained in Ref. \cite{Aguilera-Verdugo:2020nrp}.

\begin{figure}[thb!]
\centering
\includegraphics[scale=0.8]{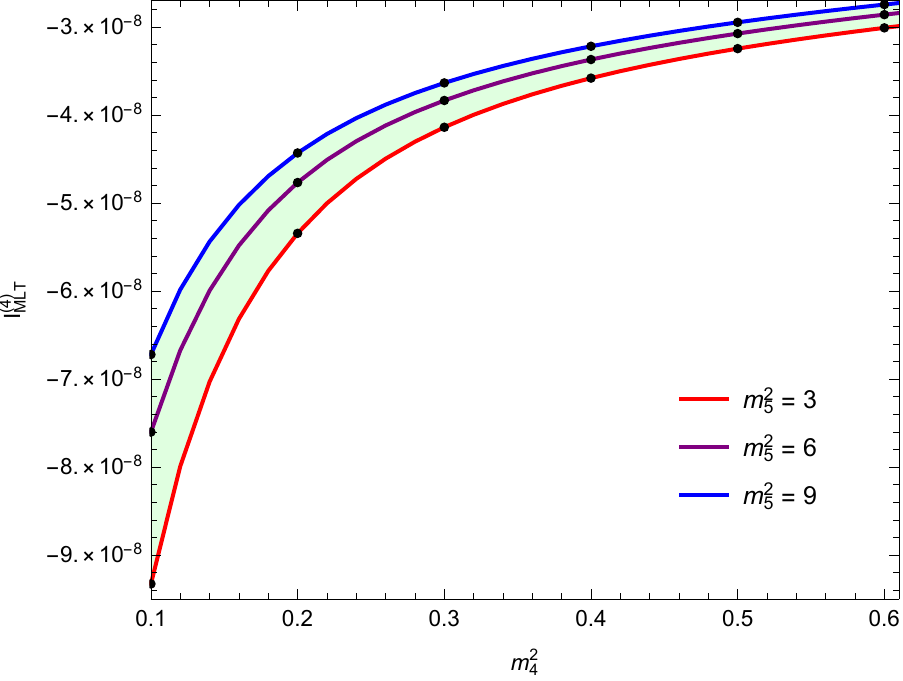}
\includegraphics[scale=0.8]{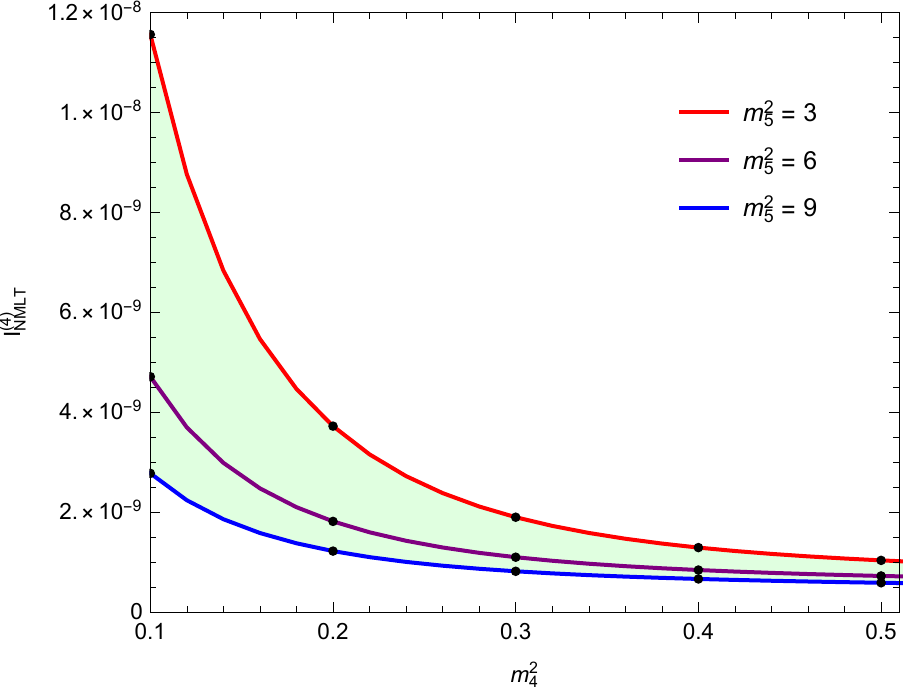}
\includegraphics[scale=0.8]{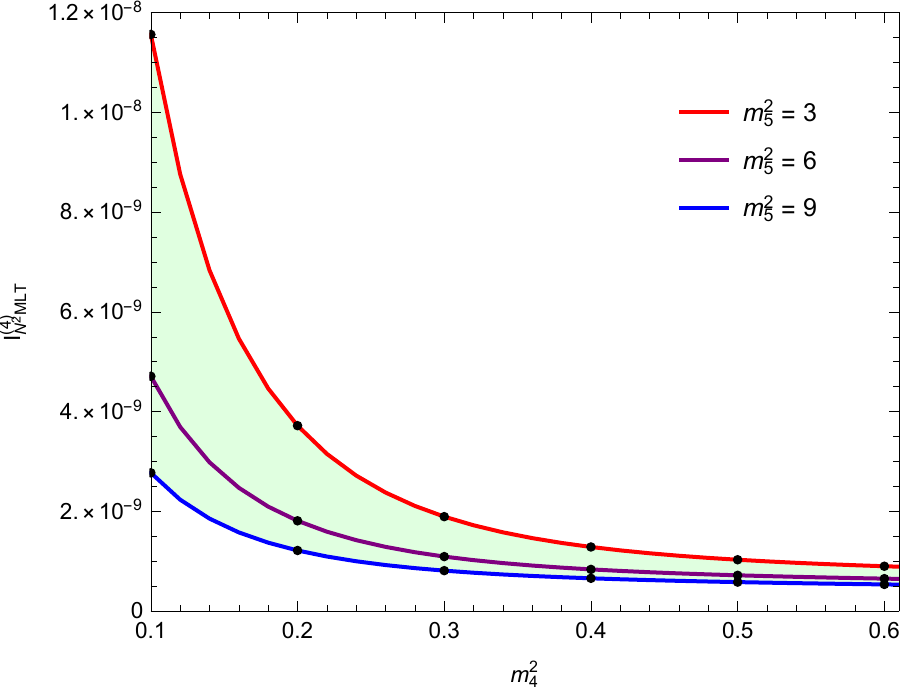}
\caption{Four-dimensional MLT, NMLT and N$^2$MLT at four loops, as a function of the internal masses $m_4^2$ and $m_5^2$. Whilst the solid lines corresponds to the results obtained within the LTD formalism, the dots are the numerical results obtained with \Fiesta.}
\label{fig:4dintegrals}
\end{figure}

\subsection{Numerical implementations}
\label{ssec:numericaleval}
Finally, we would like to highlight that the causal representation of multi-loop multi-leg Feynman integrals and scattering amplitudes leads to a very smooth numerical integration. In Ref. \cite{Aguilera-Verdugo:2020kzc}, we studied several examples and compared the performance of the LTD-inspired approach with other standard frameworks (such as \Fiesta~and \SecDec). We considered the generic scalar amplitude
\begin{align}
 {\cal A}_{\text{N}^{k-1}\text{MLT}}^{\left(L\right)} \big(1^{2},2^{2},\ldots, L^{2}&, L+1, \dots, L+k \big) \nn \\
&\,=\prod_{i=1}^{L}\frac{\partial}{\partial(q_{i,0}^{\left(+\right)})^{2}} {\cal A}_{\text{N}^{k-1}\text{MLT}}^{\left(L\right)}\left(1,2,\hdots,L+1,
\ldots, L+k \right)\,,
\label{eq:EJEMPLONUMERICO}
\end{align}
with $k=\{0,1,2\}$, at three and four loops, and changing the dimensionality of the space-time from $d=2$ to $d=4$. Here, we applied the differential operator detailed in Sec. \ref{ssec:BeyondSimple} in order to raise the powers of the denominators and achieve integrability in the UV region. In Fig. \ref{fig:4dintegrals}, we show a few examples of 
\Eq{eq:EJEMPLONUMERICO} for $d=4$, doing a scan in $m_4^2$, by fixing $m_5^2$.

The inclusion of arbitrary masses does not introduce any additional complication within the LTD-based framework. Also, only $(d-1)$ integrations for each loop are needed, whilst any methodology based on the traditional Feynman parametrisation approaches scales with the number of propagators. Moreover, we explicitly checked that the absence of non-causal singularities leads to a very efficient integration. The technology to perform these calculations was included in the \textsc{Wolfram Mathematica} package \textsc{Lotty}, recently published by one of the authors \cite{TorresBobadilla:2021dkq}.

%% file: draft_Selo.tex

The multi-loop topologies that appear for the first time at four loops are characterised by multi-loop diagrams with $L+4$ and $L+5$ sets of propagators which correspond to the next-to-next-to-next-to maximal loop topology (N$^3$MLT) and next-to-next-to-next-to-next-to maximal loop topology (N$^4$MLT). In fact, N$^4$MLT naturally includes all N$^{k-1}$MLT configurations, with $k\le 4$.

\begin{figure}[t!]
\includegraphics[scale=0.8]{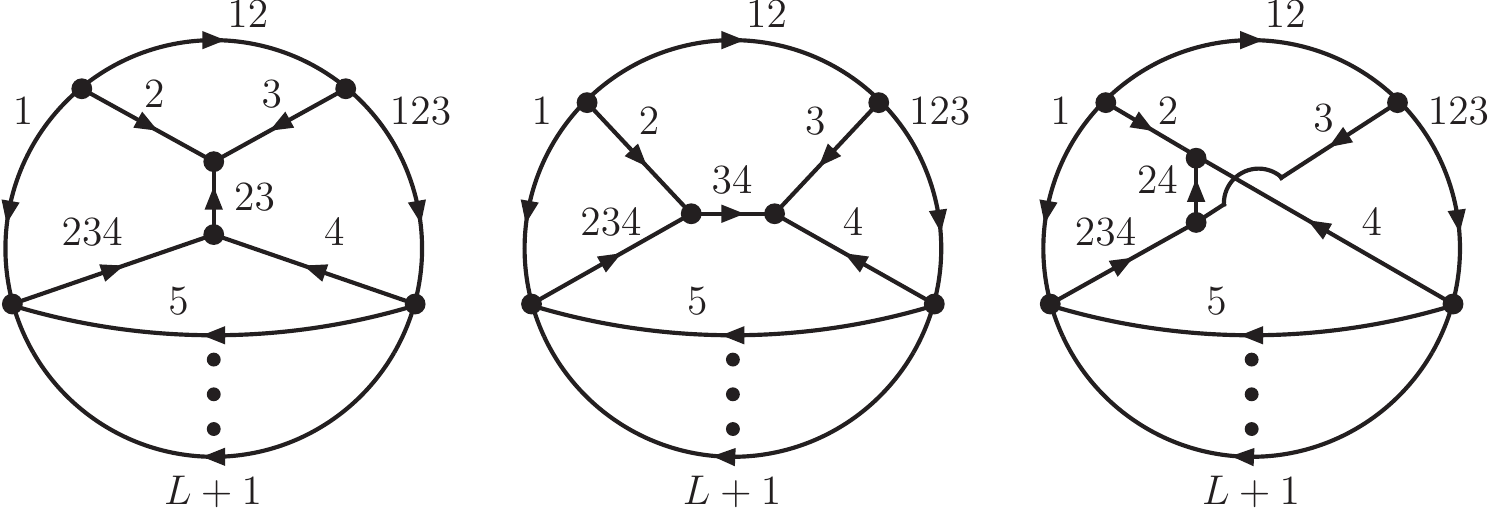}
\centering
\caption{Diagrams of the $\rm N^4MLT$ family. 
The diagram on the l.h.s. corresponds to the $t$ channel, the diagram on the center is the $s$ channel 
and the diagram on the r.h.s. corresponds to the $u$ channel. External particles are not shown. 
\label{stu-diagrams}}
\end{figure}

The N$^4$MLT family is represented with three main topologies, which were checked with \verb|QGRAF|~\cite{Nogueira:1991ex} and are shown in Fig.~\ref{stu-diagrams}. Based in the similarity of these topologies with the insertion of a four-point subamplitude with trivalent vertices in a larger topology, a unified description is proposed. The three N$^4$MLT topologies are interpreted as the $t$-, $s$- and $u$-kinematic channels, respectively, of a {\it universal topology}.

The three topologies contain $L+4$ common sets of propagators, and one extra set which is different for each of them.
Each of the first $L$ sets depends on one characteristic loop momentum $\ell_s$, the remaining four common sets are established as linear combinations of the loop momenta. The extra sets are the distinctive key to each of the channels in {\it the universal topology} where the momenta of their propagators is identified as different linear combinations of $\ell_2$, $\ell_3$ and $\ell_4$.

To assemble the three N$^4$MLT channels into a single topology, a current $J$ is defined as the union of three sets that characterise each channel,
\beq
J \equiv 23 \cup 34 \cup 24~.
\eeq
Due to momentum conservation, the three subsets do not contribute to the same individual Feynman diagram but they all contribute at amplitude level. 
Based on the described framework, the Feynman representation of the N$^4$MLT universal topology is expressed as
\begin{align}
\mathcal{A}_{\rm N^4MLT}^{(L)} (1, \ldots, L+1, 12, 123, 234, J) = \int_{\ell_1,\dots ,\ell_L}\mathcal{A}_F^{(L)}(1,\dots, L+1,12,123,234,J)~.
\end{align}
\begin{figure}[t!]
\includegraphics[scale=0.8]{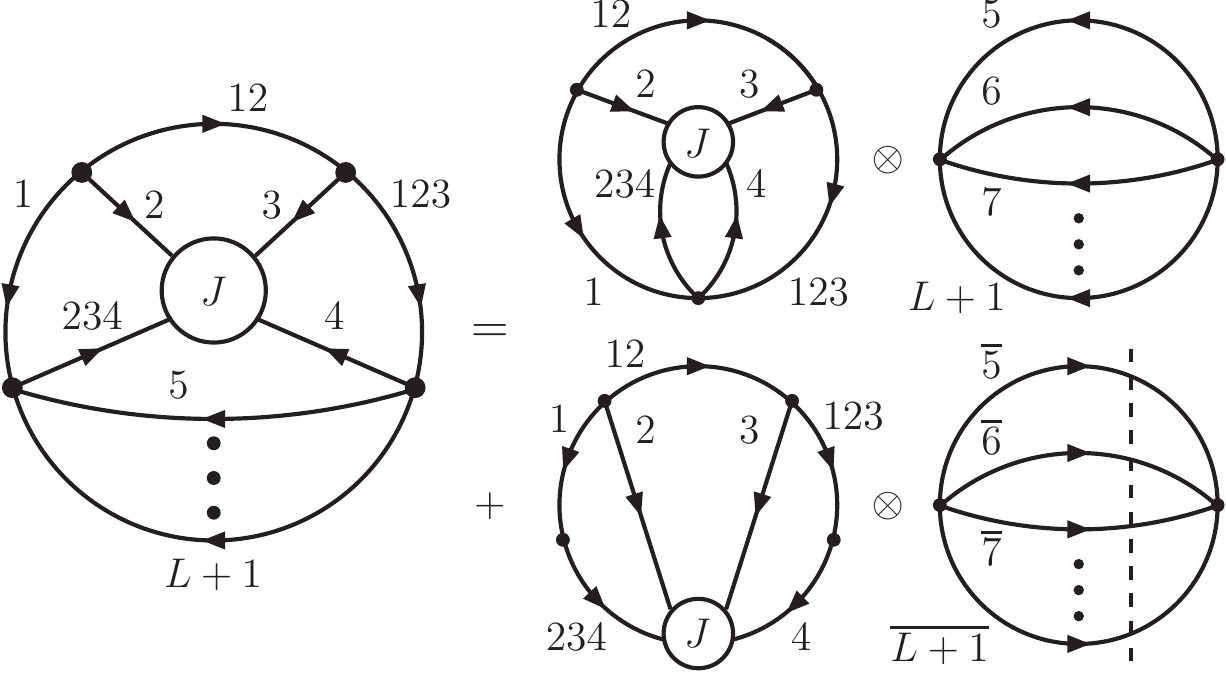}
\centering
\caption{Diagrammatic representation for the factorised opening of the multiloop 
N$^4$MLT {\it universal topology}. Only the on-shell cut of the last MLT-like subtopology 
with reversed momentum flow is shown.
\label{fig:master}}
\end{figure}
The dual opening of this topology fulfills a factorisation identity
similar to those presented in Ref.~\cite{Verdugo:2020kzh} for NMLT and N$^2$MLT
\begin{align}
\label{eq:master}
{\cal A}^{(L)}_{\rm N^4MLT} & (1, \ldots, L+1, 12, 123, 234, J) \nn \\
& =  {\cal A}^{(4)}_{\rm N^4MLT} (1, 2, 3, 4, 12, 123, 234, J) \otimes {\cal A}_{\rm MLT}^{(L-4)} (5, \dots, L+1) \nn \\
& + {\cal A}^{(3)}_{\rm N^2MLT} (1\cup 234, 2, 3, 4\cup 123, 12, J) \otimes {\cal A}_{\rm MLT}^{(L-3)} (\overline 5, \dots, \overline{L+1} )~, 
\end{align}
also valid regardless of the internal configuration.
This factorisation identity is depicted in Fig.~\ref{fig:master} and is called the universal identity in view of the fact that it is the only master expression required to open to nondisjoint trees any scattering amplitude of up to four loops. It also enables to infer the causal structure of the complete topology by exploring the causal behaviour of its subtopologies. 

The term $\mathcal{A}^{(4)}_{\rm N^4MLT}$ on the r.h.s. of \Eq{eq:master} considers all possible configurations with four on-shell propagators in the sets $\{1, 2, 3 ,4,12,123,234,J \}$, 
while ${\cal A}^{(3)}_{\rm N^2MLT}$ in the second term
assumes three on-shell conditions.
The terms ${\cal A}_{\rm MLT}^{(L-4)} (5, \dots, L+1)$ and ${\cal A}_{\rm MLT}^{(L-3)} (\overline 5, \dots, \overline {L+1})$ are computed according to  Ref.~\cite{Verdugo:2020kzh,Aguilera-Verdugo:2020nrp}.

The four-loop subtopology in \Eq{eq:master} is opened through a factorisation identity which is written in terms of known subtopologies, 
\begin{align}
\label{eq:fourloop}
\mathcal{A}_{\rm N^4MLT}^{(4)} (1, 2, 3, 4, 12, 123, 234, J) 
& = \mathcal{A}_{\rm{N^2MLT}}^{(4)}(1, 2, 3, 4, 12, 123, 234) \otimes \mathcal{A}^{(0)}(J)  \nn \\ 
& +\sum_{{\bf s}\in J}  \mathcal{A}_D^{(4)}(1, 2, 3, 4, 12, 123, 234, {\bf s}) \, .
\end{align}
Exhibiting a similar structure to \Eq{eq:fourloop}, the three-loop subtopology in \Eq{eq:master} is given by  
\begin{align}
\label{eq:threeloop}
{\cal A}^{(3)}_{\rm N^2MLT} (1\cup 234, 2, 3, 4\cup 123, 12, J) 
& = \mathcal{A}_{\rm{NMLT}}^{(3)}(1\cup234, 2, 3, 4\cup 123, 12)\otimes \mathcal{A}^{(0)}(J) \nn \\ 
& + \sum_{{\bf s}\in J}  \mathcal{A}_D^{(3)}(1, 2, 3, 4, 12, 123, 234, {\bf s})  \, ,
\end{align} 
where the bold symbol ${\bf s}$ is used to indicate that these contributions are those containing on-shell propagators in the $J$-sets.

\begin{figure}[t!]
\includegraphics[scale=0.8]{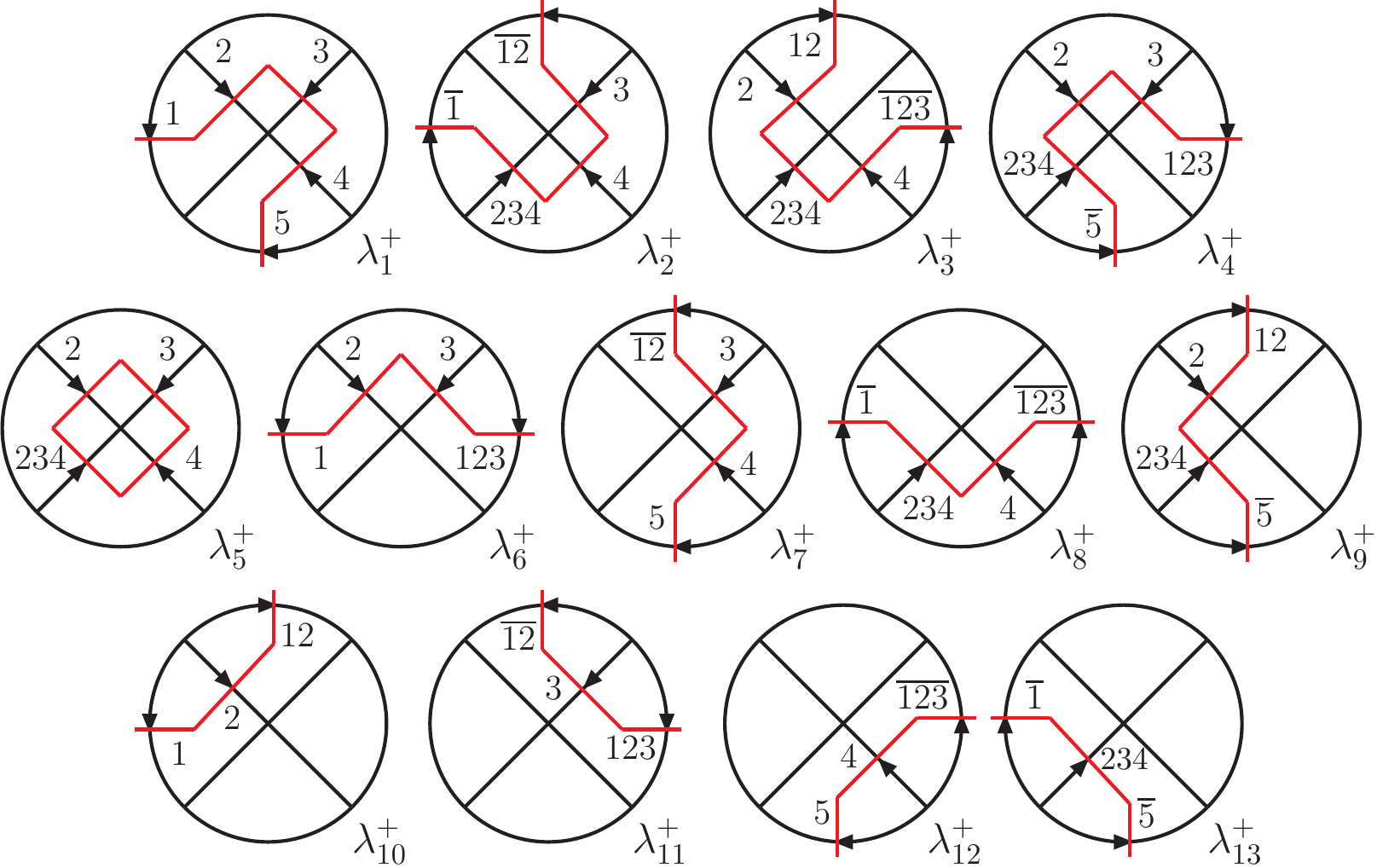}
\centering
\caption{Causal configurations of the ${\rm N^3MLT}$ topology. The set number $5$ accounts for all the propagators in the sets $5$ to $L+1$.
\label{fig:N3MLTcausal}}
\end{figure}
 
The first term on the r.h.s. of \Eq{eq:fourloop} and \Eq{eq:threeloop} consists of a four-loop N$^2$MLT and three-loop NMLT subtopology respectively. These terms are describing dual trees where all the propagators with momenta in $J$ remain off shell. The second term on the r.h.s. of \Eq{eq:fourloop} and \Eq{eq:threeloop} collects contributions that characterise the $s$, $t$ or $u$ channel shown in Fig.~\ref{stu-diagrams}. To obtain these contributions a propagator in either set $23$, $34$ or $24$ are set on shell.

The explicit expressions for the $s$, $t$ and $u$ channel arising from \Eq{eq:fourloop} and \Eq{eq:threeloop} are presented in Ref. \cite{Ramirez-Uribe:2020hes} where all the results are consistent with the absence of disjoint trees. Repeated propagators from self-energy insertions were treated as single propagators raised to specific powers and are not considered to generate disjoint trees when the repeated propagator is set on shell.

Notice that the number of trees in the LTD forest can also be computed 
through the combinatorial exercise of selecting, from the full list of sets, all possible subsets of $L$ elements that cannot generate disjoint trees. Nevertheless, the momentum flows of the on-shell propagators can only be determined through the nested residues.
Also, it is important to mention that the number of terms for a given N$^{k-1}$MLT topology in \Eq{eq:master} scales with the number of loops and linearly with the number of propagators per loop set, but the sum over residues, equivalently over internal propagators, is implicitly accounted in this expression.

\subsection{Causal representations}
\label{ssec:SELOcausal}
To confirm the causal conjecture for the N$^4$MLT family, the strategy proposed in Ref. \cite{Aguilera-Verdugo:2020kzc} is applied to the multi-loop N$^3$MLT, $t$, $s$ and $u$ channels. Each topology considers a configuration with one internal propagator in each loop set, four external momenta for N$^3$MLT and six external particles for $t$, $s$ and $u$ channels.

As a first step, the LTD representation is obtained for each of the selected topologies through the universal N$^4$MLT expression in \Eq{eq:master}. After computing the nested residues and adding them all together a causal expression is found. The integrand of the causal dual representation reads in terms of on-shell energies, the energy components of the linear combination and causal denominators. Let us remember and emphasise that these causal denominators are constructed from sums of on-shell energies exclusively, and they represent potential singular configuration. 

\begin{figure}[t!]
\includegraphics[scale=0.7]{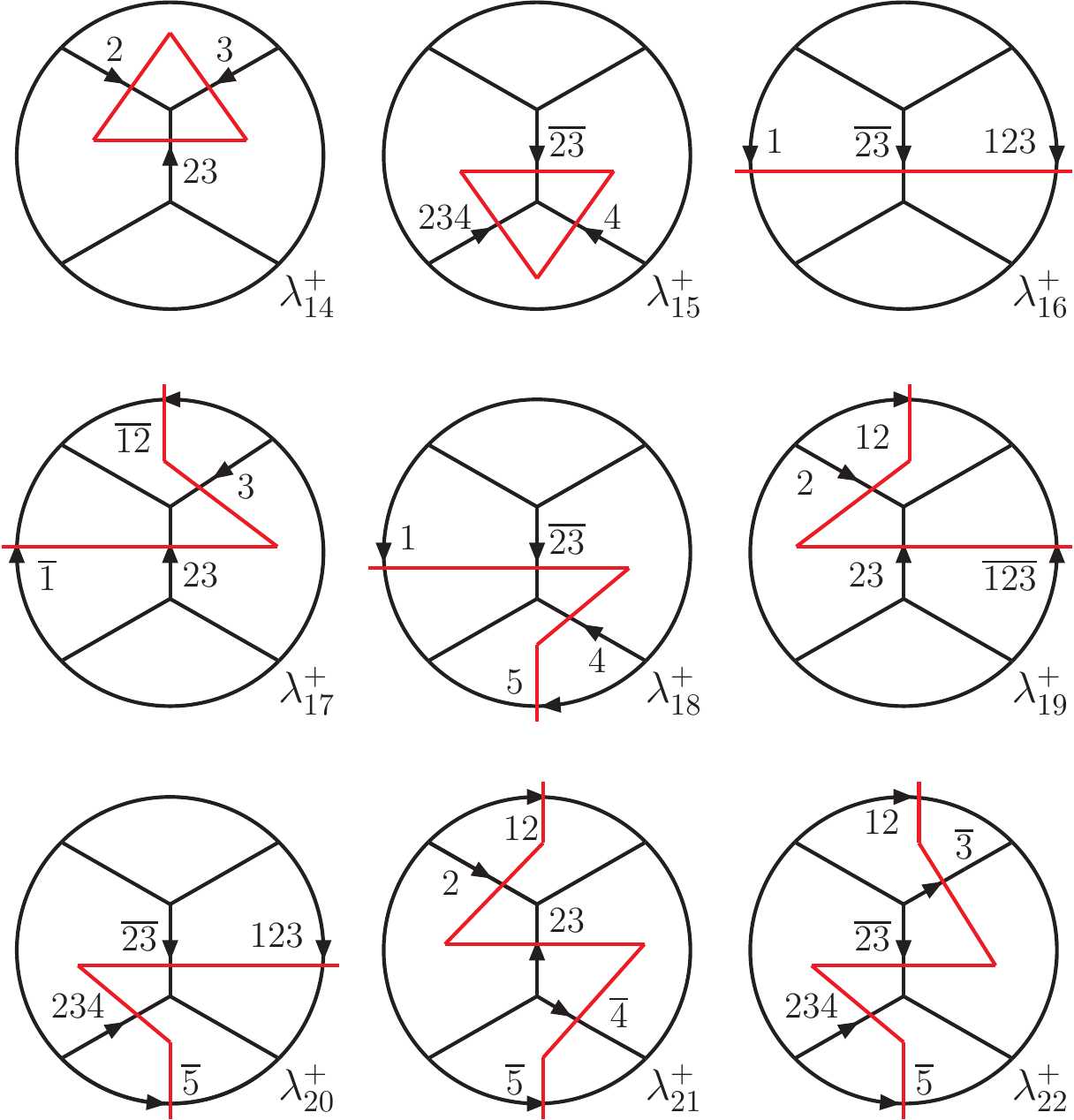}
\centering
\caption{Extra causal configurations of the $t$-channel of the ${\rm N^4MLT}$ topology. 
\label{fig:N4MLTtcausal}}
\end{figure}

The results given by the straightforward application of LTD leads directly to a manifestly causal expression, however, the resulting numerator is a lengthy polynomial in the on-shell and external energies \cite{Ramirez-Uribe:2020hes}. Therefore, a way to get a more suitable causal expression is to reinterpret it in terms of a number of entangled thresholds equal to the difference between the number of propagators and the number of loops by, e.g., analytical reconstruction from numerical evaluation over finite fields~\cite{vonManteuffel:2014ixa,Peraro:2016wsq,Peraro:2019svx} as defined in Ref.~\cite{Aguilera-Verdugo:2020kzc}. 

The discussion was stated only for scalar integrals given that they fully encode all the compatible causal combinations. In general, tensor reduction commutes with LTD and can be used to deal with tensor integrals. Another feature exploited is that external momenta attached to interaction vertices that connect different loop sets do not alter the number of internal propagators and therefore the complexity of the causal representation. Therefore, the full causal expressions with external momenta can be deduced from the causal representation of the vacuum configuration by matching the momentum flows of the entangled thresholds.

Starting with the multi-loop N$^3$MLT, there are thirteen causal denominators which are depicted in Fig.~\ref{fig:N3MLTcausal}. The analytically reconstruction was done by matching all combinations of four thresholds that are causally compatible to each other.

Going forward to the causal representation of the N$^4$MLT family,  we have to consider all the entangled configurations with the presence of five causal thresholds. The $t$ channel depends on the causal denominators already defined for the N$^3$MLT configuration and additional nine extra causal denominators that depend on $\qon{23}$ where the corresponding configurations are shown in Fig.~\ref{fig:N4MLTtcausal}.

\begin{figure}[t!]
\includegraphics[scale=0.7]{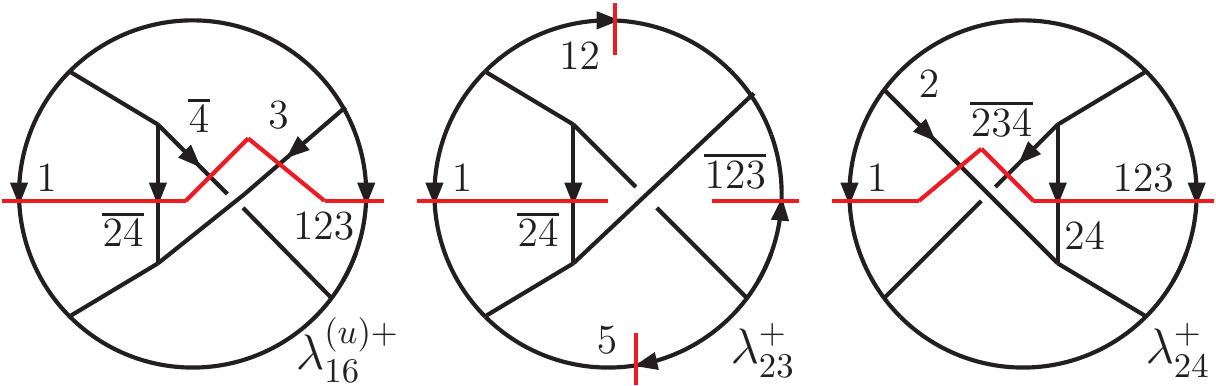}
\centering
\caption{Extra causal configurations of the $u$-channel of the ${\rm N^4MLT}$ topology due to nonplanarity.
\label{fig:N4MLTucausal}}
\end{figure}

The $s$-channel can be obtained through the structure of the $t$-channel. The causal denominators are obtained just by a clockwise rotation of the $t$-channel and therefore by a permutation of the arguments of the causal denominators that are channel specific.
Similar to the $s$-channel, the $u$-channel is obtained from the $t$-channel through the substitution $23\to 24$ and by the exchange $3\leftrightarrow 4$ or $2\leftrightarrow 234$  ($123$ remains invariant). 
There are, however, three new configurations that arise because the $u$-channel is nonplanar. These new configurations
are shown in  Fig.~\ref{fig:N4MLTucausal}.

One advantage of the causal representation is that the number of terms for a given N$^{k-1}$MLT topology is independent of the number of loops in the causal representation 
but requires to specify additional causal thresholds and additional causal entanglements when more internal propagators are considered. Furthermore, the most significant advantage of the causal representation with respect to the LTD representation stems from the core difference between them, the presence or absence of noncausal singularities. The straightforward application of the nested residue generates multiple threshold singularities, nevertheless, with a clever analytical rearrangement, the absence of noncausal singularities is achieved and leads to a causal representation which is more efficient and stable numerically in all the integration domain \cite{Ramirez-Uribe:2020hes}.

%% file: draft_GS.tex
The applications of the LTD-inspired methods have been spreading very fast in the last years. In particular, it turned out to be an excellent tool to unveil the structure of causal singularities of multi-loop multi-leg Feynman integrals and scattering amplitudes. There have been very recent findings regarding general all-loop formulae to describe any N$^k$MLT amplitude, by using clever algebraic relations among them \cite{Bobadilla:2021rmu}. This computational technology has been implemented in the package \textsc{Lotty} \cite{TorresBobadilla:2021dkq}, which allows to automatically obtain the causal representation of multi-loop multi-loop Feynman integrals and scattering amplitudes.

\begin{figure}[ht]
    \centering
    \includegraphics[width=0.50\textwidth]{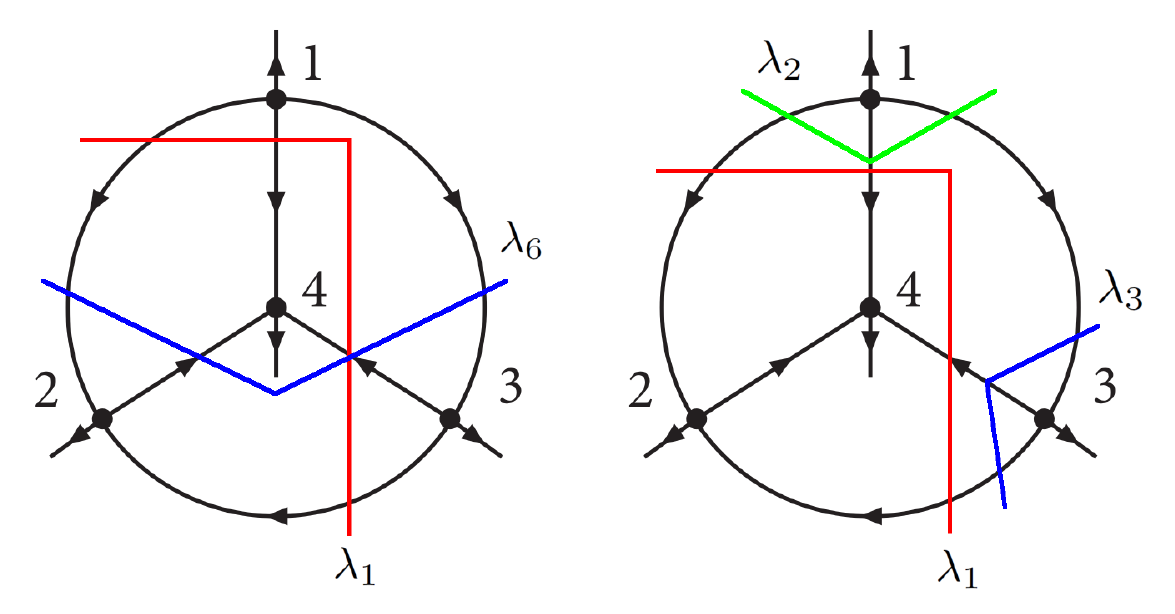}
    \caption{Example of two incompatible causal propagators with non-disjoint sets of cusps (left) and incompatible momenta orientation (right).
    }
    \label{fig:geometrical1}
\end{figure}

On the other hand, investigations inquiring on the geometrical aspects of multi-loop diagrams were performed. In Ref. \cite{Sborlini:2021owe}, we used concepts from graph theory to describe all the possible causal propagators and the allowed entangled thresholds associated to a given diagram. Describing diagrams in terms of cusps (i.e. interaction vertices) and edges (i.e. sets of propagators connecting two fixed vertices) was also proposed in Ref. \cite{Bobadilla:2021rmu}. This description leads to the \emph{cusp matrix}, a geometrical object that allows one to completely characterise the causal structure of a given diagram. Explicitly, we realised that the causal propagators are associated to all the possible connected binary partitions of cusps. This has a very nice connection to a generalised geometrical interpretation of the Cutkosky's rules \cite{Cutkosky:1960sp}. Also, we implemented a computational algorithm that benefits from graph theory tools to efficiently obtain all the causal propagators. We tested it against the direct nested residue calculation and subsequent identification of $\lambda$'s: the geometric algorithm was, at least, a factor ten faster.

The information contained in the cusp matrix allow us to exactly reconstruct all the possible entangled thresholds by imposing geometrical selection criteria. These can be summarised in the following list:
\begin{enumerate}
    \item \emph{All the lines are cut}: Each possible entangled combination of causal denominators must involve the on-shell energies of all the propagators.
    \item \emph{Absence of crossings}: Causal denominators are associated to connected partitions of vertices. Only those partitions involving disjoint sets of vertices, or those that are strictly included, can be successfully entangled. Alternatively, if we associate a line connecting the cut edges to each $\lambda$, then two $\lambda$'s can be entangled if their lines do no cross.
    \item \emph{Consistent flux orientation}: A set of causal propagators is compatible (or can be entangled) if the associated binary partitions can consistently be oriented, i.e. if all the internal lines contribute to the entangled cuts with the same orientation.
    \item \emph{Causal propagator orientation}: When external particles are present, the relative signs among aligned on-shell energies (i.e. $\sum q_{i,0}^{(+)}$) and the energy component of the external momenta is determined by the \emph{orientation matrix}, which can be built from the cusp matrix.
\end{enumerate}
In Fig. \ref{fig:geometrical1}, we show examples of causal propagators which can not be entangled for the Mercedes-Benz diagram (i.e. a particular case of N$^2$MLT or a four-cusp topology). On the left, the criteria 2 is not fulfilled, since the associated partitions of cusps are not disjoint. On the right, we consider three causal propagators which can not be consistently oriented (criteria 3): the edge connecting the cusps 1 and 3 does not allow a compatible orientation. 

Thus, given a diagram, determining its cusp matrix and imposing the selection criteria 1-4, we can obtain a formula describing its causal structure. In accordance with the results presented in Ref. \cite{Bobadilla:2021rmu}, we found that
\begin{align}
 {\cal A}_{N}^{\left(L\right)}(1,\ldots,L+k)=\sum_{\sigma \in \Sigma}  \, \int_{\vec{\ell}_{1},\cdots,\vec{\ell}_{L}}\frac{{\cal N}_{\sigma}(\{q_{r,0}^{(+)}\},\{p_{j,0}\})}{x_{L+k}} \, \times \prod_{i=1}^{k} \frac{1}{-\lambda_{\sigma(i)}}\, + \, (\sigma \leftrightarrow \bar{\sigma})  \, ,
\label{eq:MasterCausalFormula}
\end{align}
describes the causal structure of any multi-loop multi-leg Feynman diagram or collection of Feynman diagrams with topological complexity $k-1$. The set $\Sigma$ contains all the subsets of products of $k$ causal propagators $\lambda_i^{\pm}$ fulfilling the selection criteria 1-4, and ${\cal N}_\sigma$ is given by the application of an operator depending on the subset of $\sigma$. Thus, this result confirms that multi-loop multi-leg amplitudes can be purely described in terms of cusps and edges, regardless the number of loops or external lines.

%% file: draft_Roger.tex
The standard computation of accurate predictions at Next-to-Leading Order (NLO) in QFT requires to deal with non trivial integrands. These integrands carry multiple singularities due to the low (IR) and high (UV) energy regimes. The cancellation of those divergent quantities, typically occurs at integral-level. By considering the Kinoshita-Lee-Nauenberg (KLN) theorem~\cite{Kinoshita:1962ur,Lee:1964is}, soft and collinear divergences (IR singularities) are removed, and by adding suitable ultraviolet counter-terms, UV singularities are renormalised, rendering the cross section finite. This algorithm for cancelling singularities can be extended to Next-to-Next-to-Leading order (NNLO) and beyond; however, since the complexity of the integrands is higher at higher orders, this procedure is reaching a bottleneck because cancellation of singularities must be achieved at integral level, i.e., the $\epsilon\to 0$ limit is taken once the cross-section is known in DREG. Since LTD transform loop integrals in phase-space integrals, among other properties discussed in this document, the cancellation of singularities could occurs at integrand level. Hence, based on the LTD theorem, the {\it Four Dimensional Unsubtraction}~\cite{Hernandez-Pinto:2015ysa,Sborlini:2015uia,Sborlini:2016fcj,Sborlini:2016gbr,Sborlini:2016hat} (FDU) method presents a new paradigm to compute observables in four dimensions, since the definition of cross-section is free of IR and UV singularities. The first application of the LTD at cross-section level was done in a toy model based on the simplest $\phi^3$ theory. In Ref.~\cite{Hernandez-Pinto:2015ysa} it was shown for the first time that LTD is powerful to build physical observables in four dimensions by a proper mapping between real and virtual kinematical variables.

\subsection{Local cancellation of infrared singularities within FDU}
\label{ssec:CancelaIR}
Let us start the discussion of the FDU approach by analysing the cancellation of soft and collinear divergences. The simplest scenario is the decay process $1\to n$ where the Born level cross-section is given by,
\begin{align}
    \sigma^{(0)}=\int {\rm dPS}^{1\to n} \, \vert \mathcal{M}_n^{(0)}\vert^2 \,  \mathcal{S}_0(\{p_i\}) \, ,
\end{align}
with $\vert\mathcal{M}_n^{(0)}\vert^2$ the LO contribution and $\mathcal{S}_0$ is the IR-safe measure function. The virtual correction to the LO cross-section is computed as
\begin{align}
    \sigma^{(1)}_V=\int {\rm dPS}^{1\to n} \int_{\ell}2{\rm Re}\, \langle  \mathcal{M}_n^{(0)} \vert \mathcal{M}_n^{(1)}\rangle \, \mathcal{S}_0(\{p_i\}) \, ,
\end{align}
where $\langle  \mathcal{M}_n^{(0)} \vert \mathcal{M}_n^{(1)}\rangle$ represents the interference between the Born level and one-loop amplitudes. It is important to emphasise at this point that FDU requires also to maintain self-energy contributions because they contain both IR and UV divergences that will contribute to the full cancellation of singularities. The IR singularities from the virtual component are cancelled against real radiation contributions. 

For the sake of simplicity, let us consider that the process under study only has final-state radiation singularities. Hence, the real contribution given by
\begin{align}
    \sigma^{(1)}_R=\int {\rm dPS}^{1\to n+1} \, \vert \mathcal{M}_{n+1}^{(0)}\vert^2 \,  \mathcal{S}_1(\{p'_i\}) \, ,
\end{align}
contains all Feynman diagrams with one extra particle in the final state in $\mathcal{M}_{n+1}^{(0)}$ and $\mathcal{S}_1$ is the measure function for $n+1$ particles. It is important to recall at this moment that primed variables are used to describe the momenta of real matrix elements while unprimed variables are labelling Born and loop level momenta.

The FDU starts with the analysis of the virtual cross-section which can be decomposed as a sum of dual contributions as,
\begin{align}
    \sigma^{(1)}_{V} & =\int {\rm dPS}^{1\to n} \, \sum_{i=1}^N \int_{\vec{\ell}} I_i(q_i) \, \mathcal{S}_0(\{ p_i \}) \nn \\
    &\equiv \int {\rm dPS}^{1\to n} \,\sum_{i=1}^N \sigma_{D,i}^{(1)} \, ,
\end{align}
with $I_i(q_i)$ the $N$ dual integrals that arise from the direct application of the LTD theorem. We notice that each $I_i(q_i)$ has set one internal particle on shell and it is characterised by $q_i$. Then, the loop integral has been transformed to phase-space integrals, therefore each dual integral behaves as the contribution of one real particle emitted. In order to achieve the complete cancellation of IR singularities, each $\sigma_{D,i}^{(1)}$ must be paired, at integrand level with a corresponding real radiation twin component. The real radiation is sliced as,
\begin{align}
    \sigma^{(1)}_{R,i}=\int {\rm dPS}^{1\to n+1} \, d\sigma_R^{(1)} \, \mathcal{R}_i \, ,
\end{align}
where $\mathcal{R}_i$ represents a partition of the full integral, such that $\sum_i \sigma_{R,i}^{(1)}=\sigma_R^{(1)}$ and considering that only one IR divergent configuration is allowed in each $\mathcal{R}_i$. Finally, in order to merge both integrals, a mapping between the kinematical variables is implemented. On the one hand, each partition in $\sigma_{D,i}^{(1)}$ contains $n$ external momenta plus one extra integration variable corresponding to $q_i$, therefore, each $\sigma_{D,i}^{(1)}$ is mapped into its real sliced contribution as,
\begin{align}
    \mathcal{T}_i(\{ p_1,\dots , p_n, q_i\}) \to \{ p'_1,\dots , p'_{n+1}\} \, ,
\end{align}
where $\mathcal{T}_i$ is a bijective transformation among real and virtual variables. The momentum mapping between variables is analogous to the one used by the dipole method~\cite{Catani:1996jh,Catani:1996vz}, since the singularities are associated to soft emissions or the double collinear limit~\cite{Kosower:1999rx,Sborlini:2013jba}. These singular IR configurations are associated to specific contributions in the virtual part, by selecting two massless partons per partition. On one hand the {\it spectator} and on the other hand the {\it emitter}. Hence, the four momenta of the emitter and spectator in addition with the loop three-momentum are used to reconstruct the kinematical phase-space of the real emission cross-section where there is a similar configuration, i.e., an emitter decaying into two partons in a soft or collinear regime. Explicitly, if $p_i$ is the momentum of the final-state emitter, $q_i$ the internal on-shell momentum prior the emitter and $p_j$ is the momentum of the final-state spectator, we apply the momentum mapping
\begin{align}
    p_r^{\prime\mu} &= q_i^{\mu} \, ,\nn \\
    p_i^{\prime\mu} & = p_i^{\mu}-q_i^{\mu}+\alpha_i \, p_j^{\mu} \, , \nn \\
    p_j^{\prime\mu} & = (1-\alpha_i)  \, p_j^{\mu} \, , \nn  \\
    p_k^{\prime\mu} & =   \, p_k^{\mu} \, , \qquad \qquad \qquad k\neq i,j \, ,
\end{align}
with $p_r'$ the momentum of the extra radiation of the process and, in the massless scenario, $\alpha_i = (q_i-p_i)^2/(2p_j\cdot(q_i-p_i))$. Furthermore, this mapping preserves momentum conservation since $p_i + p_j +\sum_{k\neq i, j} p_k =p_i' + p_j' +p_r' +\sum_{k\neq i, j} p'_k$ is fulfilled. A similar mapping is used when considering massive particles and it was shown in Ref~\cite{Sborlini:2016hat}. The extension of this formalism to higher-orders (i.e. NNLO and beyond) will require additional kinematical transformations which take care of the singular behaviour of scattering amplitudes in the multiple-collinear limit \cite{Catani:1999ss,Catani:2003vu,Sborlini:2014eib,Sborlini:2014mpa,Sborlini:2014kla,DelDuca:2019ggv,DelDuca:2020vst}.

\subsection{Self-energy insertions and renormalisation}
\label{ssec:SelfenergyFDU}
Before moving forward with IR cancellation, let us review the local renormalisation of UV singularities. The standard cancellation of UV singularities requires the renormalisation of field wave-functions and couplings. In the FDU formalism, this feature is obtained through the construction of local UV counter-terms. At one-loop, the scenario where massless and massive particles are propagating in the loop has been studied~\cite{Sborlini:2016gbr,Sborlini:2016hat}. Remarkably, it was shown that a smooth transition between massless and massive renormalisation constants takes place and UV singularities are well understood. Let us highlight a crucial difference between the standard renormalisation constant in DREG and LTD. Wave-function renormalisation constants are obtained from self-energy diagrams. In particular, massless bubble diagrams in DREG are neglected in the renormalisation procedure since IR and UV divergences are considered as equal and they cancel out. However, the same divergent poles in the FDU formalism contribute separately. Specifically, it means that IR singularities of loop diagrams vanish with only the IR poles of the real radiation and there are no mixed cancellation between UV and IR poles. Therefore, the remaining UV divergences must be removed when renormalisation is implemented in the FDU scheme. Hence, we stress that integrands in the FDU are separated into the IR and UV domains, and this identification is crucial to render cross-sections free of singularities in the FDU formalism. 

We analyze the construction of the renormalisation constants in the FDU framework. By using standard Feynman rules, the unintegrated massive wave-function renormalisation constant, in the Feynman gauge, at one-loop is given by,
\begin{align}\label{eq:DeltaZ2expression}
\nn \Delta Z_2(p_1;M) = -\g^2 \, C_{\rm F} \,
\int_{\ell} G_F(q_1) \, G_F(q_3) \,
\left[(d-2)\frac{q_1 \cdot p_2}{p_1 \cdot p_2} \right.
+4\, M^2 \left. \left(1- \frac{q_1 \cdot p_2}{p_1 \cdot p_2}\right)
  G_F(q_3)\right]\, . \nn \\
\end{align}
\Eq{eq:DeltaZ2expression} is the most general wave-function renormalisation constant since it includes the massless and massive case and, as it was previously mentioned, the transition to the massless case, $\Delta Z_2(p,0)$, is straightforward. The function $\Delta Z_2(p_1; M)$ contains singularities associated to the UV domain, therefore we must find the UV component of \Eq{eq:DeltaZ2expression}, $\Delta Z_2^{\rm UV}$, and subtract it in order to find a UV-free wave-function renormalisation constant, $\Delta Z_2^{\rm IR}$. The UV part is extracted by performing an expansion of the integrand around the UV propagator $G_F(q_{\rm UV}) = (q^2_{\rm UV}-\mu_{\rm UV}^2 + \imath 0)^{-1}$. In particular, for \Eq{eq:DeltaZ2expression}, it is found
\begin{align}
\nn\Delta Z_2^{\rm UV}(p_1) &= (2-d)\, \g^2 \, C_{\rm F} \,
\int_{\ell} \big[G_F(q_\text{UV})\big]^2 \,
  \left(1+\frac{q_\text{UV} \cdot p_2}{p_1 \cdot p_2}\right) 
\Big[1\!-\!G_F(q_\text{UV})(2\, q_\text{UV} \cdot p_1 + \mu^2_\text{UV})\Big] \, . \\
\label{eq:ParteUV}
\end{align}
Hence, we define $\Delta Z_2^{\rm IR}$ as
\begin{align}
\Delta Z_2^{\text{IR}} = \Delta Z_2 - \Delta Z_2^{\text{UV}}\, ,
\end{align}
which is free of UV singularities and the IR singularities poles are those needed to remove the remaining singularities from the virtual and real mapping of the cross-section. 

To remove UV singularities it is important to build proper UV counter-terms that can be extracted by the direct study of the UV properties of physical amplitudes. After the combination of real and virtual matrix elements and renormalisation, the remaining amplitude has only UV singularities, in the following, we consider generic amplitudes with only UV singularities. At two loops, a generic two loop amplitude can be written as,
\begin{align}
    \mathcal{A}^{(2)} = \int_{\ell_1}\int_{\ell_2} \mathcal{I}(\ell_1, \ell_2) \, ,
\end{align}
where the integrand is a function of the loop variables $\ell_1$ and $\ell_2$. UV divergences shall appear when the three-momenta $|\vec{\ell}_1|$ and $|\vec{\ell}_2|$ tend to infinity. In the two-loop scenario, there are three possible UV limits: {\it i)} $|\vec{\ell}_1| \to \infty$ and $|\vec{\ell}_2|$ remains fixed, {\it ii)} $|\vec{\ell}_2| \to \infty$ and $|\vec{\ell}_1|$ is fixed and, {\it iii)} $|\vec{\ell}_{1,2}| \to \infty$. In order to extract the UV behaviour of the integrand, we impose the replacement,
\begin{align}
    \mathcal{S}_{j,\rm UV} : \{ \ell_j^2\vert \ell_j \cdot k_i\} \to \{\lambda^2 q_{j, \rm UV}^2 + (1-\lambda^2)\mu_{\rm UV}^2 \vert \lambda \,  q_{j,\rm UV}\cdot k_i \} \, ,
\end{align}
for a given loop momentum $\ell_j$ and then, we expand the expression up to logarithmic order around the UV propagator. This action is represented by the $L_{\lambda}$ operator. Notice, however, that the result shall generate a finite part after integration that has to be fixed to find the right value of the integral. Therefore, the first counter-terms is computed by
\begin{align}
    \mathcal{A}_{j,\rm UV}^{(2)} = L_{\lambda} \left( \mathcal{A}^{(2)}\Big|_{\mathcal{S}_{j,\rm UV}}  \right) -d_{j,\rm UV} \, \mu_{\rm UV}^2 \int_{\ell_j} (G_F(q_{j,\rm UV}))^3 \, ,
\end{align}
with $d_{j,\rm UV}$ the fixing parameter which makes the finite part of integral to be zero in the $\overline{\rm MS}$ scheme.

The remaining divergences shall occur when both $|\vec{\ell}_1|$ and $|\vec{\ell}_2|$ approach to infinity simultaneously. Then, to build a counter-term that mimics this behaviour, the following replacement is implemented,
\begin{align}
    \mathcal{S}_{\rm UV^2} : 
    &\{ \ell_j^2\vert \ell_j \cdot \ell_k\vert \ell_j \cdot k_i\} \to
    \nn \\
    &\{\lambda^2 q_{j, \rm UV}^2 + (1-\lambda^2)\mu_{\rm UV}^2
    \vert \lambda^2 q_{j,\rm UV} \cdot q_{k, \rm UV} +(1-\lambda^2) \, \mu_{\rm UV}^2/2
    \vert \lambda \,  q_{j,\rm UV}\cdot k_i \} \,
\end{align}
on the subtracted integrand. Similarly to the previous UV counter-term, the application of the $L_{\lambda}$ operation shall produce a finite piece that has to be fixed to build properly the counter-term, $\mathcal{A}^{(2)}_{\rm UV^2}$. Explicitly, 
\begin{align}
    \mathcal{A}_{\rm UV^2}^{(2)} = L_{\lambda} \left( \left(
    \mathcal{A}^{(2)} -\sum_{j=1,2}\mathcal{A}^{(2)}_{j,\rm UV} \right) \Bigg|_{\mathcal{S_{\rm UV^2}}}
    \right) -d_{\rm UV^2} \, \mu_{\rm UV}^4 \int_{\ell_1} \int_{\ell_2} (G_F(q_{1,\rm UV}))^3(G_F(q_{12,\rm UV}))^3 \, ,
\end{align}
with $d_{\rm UV^2}$ the fixing parameter of the double UV limit. Having all UV divergences under control, the renormalised amplitude at two-loops, $\mathcal{A}_{\rm R}^{(2)}$, can be constructed by the subtraction of all UV counter-terms, such that
\begin{align}
\mathcal{A}_{\rm R}^{(2)} = \mathcal{A}^{(2)} - \mathcal{A}_{1,\rm UV}^{(2)}- \mathcal{A}_{2,\rm UV}^{(2)} - \mathcal{A}_{\rm UV^2}^{(2)} \, ,
\end{align}
is free of IR and UV singularities.

In the multi-loop scenario, multiple ultraviolet poles will appear, since all loops could tend to infinity at different {\it speed}. 

\subsection{$A^*\to q \bar{q}$ at NLO in QCD}
\label{ssec:FDUnloexample}
The FDU method has been studied in different processes, for the sake of simplicity, we recall the application of the algorithm in $A^* \to q \bar{q}$ a NLO in QCD, with $A=H, \gamma, Z$. 

\begin{figure}[t]
    \centering
    \includegraphics[width=1\textwidth]{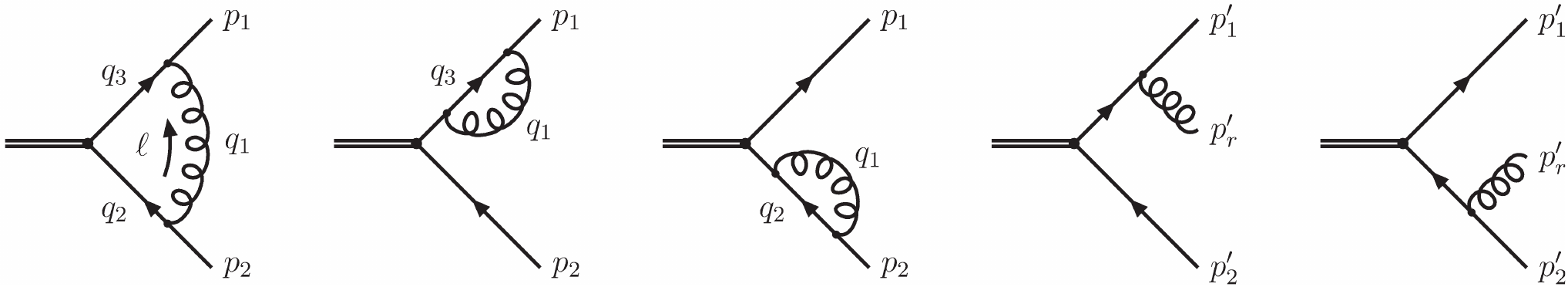}
    \caption{Momentum configuration of NLO QCD corrections to the process $A^* \to q \bar{q}(q)$, assuming that the decaying particle does not couple to gluons.
    }
    \label{fig:Aqqbar}
\end{figure}

At NLO, virtual correction $A^*(p)\to q(p_1) + \bar{q}(p_2)$ must be mapped into the real radiation process $A^*(p)\to q(p'_1) + \bar{q}(p'_2)+g(p'_r)$. As it was explained in previous subsections, we must also keep all virtual contributions since self-energy contributions are crucial to cancel out all IR singularities. In Fig. \ref{fig:Aqqbar}, we present the complete set of Feynman diagrams of the real and virtual processes.

\begin{figure}[t]
    \centering
    \includegraphics[width=0.5\textwidth]{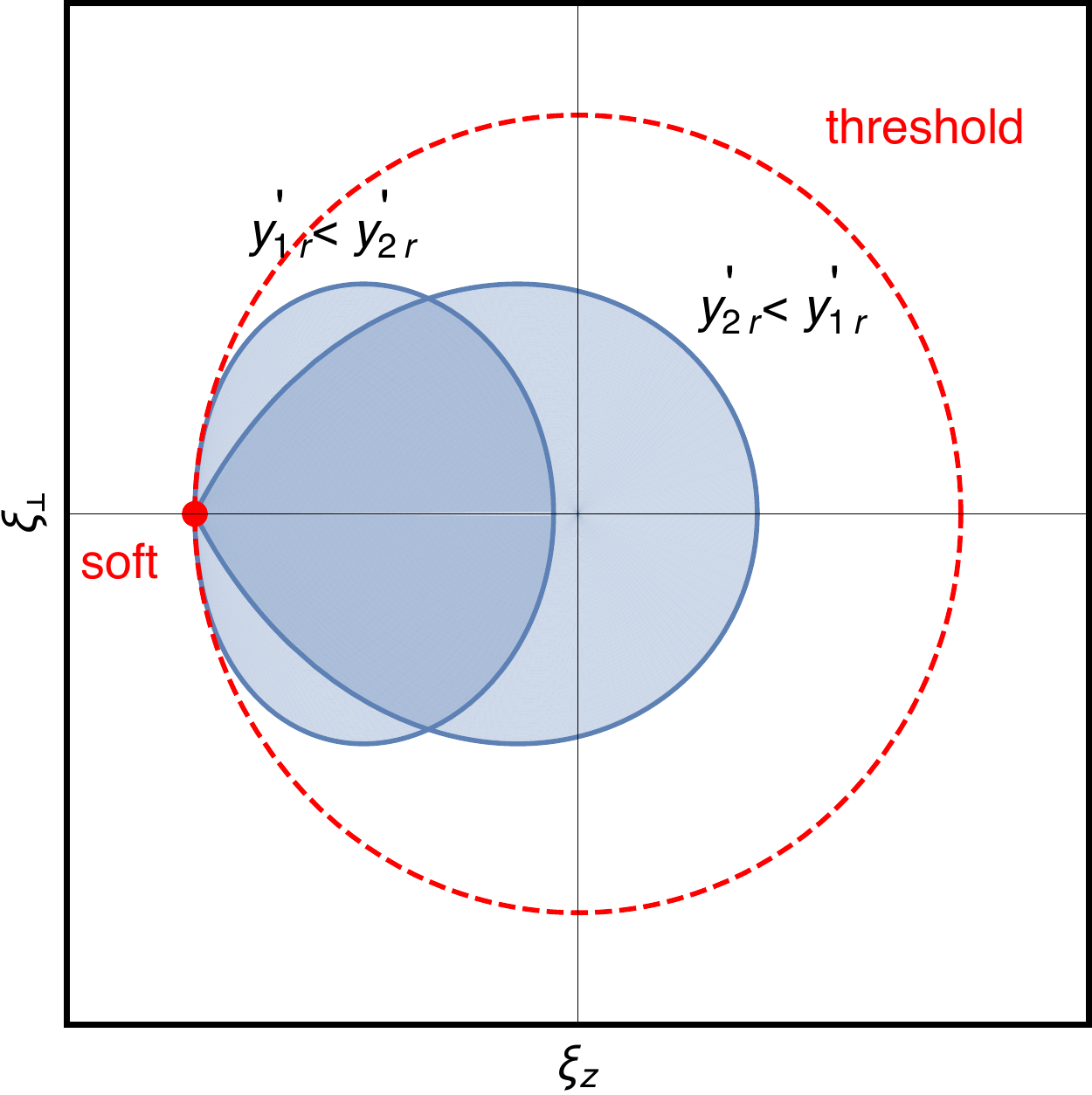}
    \caption{The dual integration regions in the loop three-momentum space, with $\xi_{\perp}=\sqrt{\xi^2_x+\xi^2_y}$.
    }
    \label{fig:dualregions}
\end{figure}

It is important to recall that in this example we consider massive quarks, i.e., $p_1^2=p_2^2=M^2$ and $p_1^{\prime 2} = p_2^{\prime 2}=M^2$ and $p_r^{\prime 2}=0$. It is convenient to write the momenta $p_1$ and $p_2$ in terms of massless four vectors, $\hat{p}_1$ and $\hat{p}_2$, such as
\begin{align}
    p_1^{\mu}= \beta_+ \hat{p}_1^{\mu}+\beta_- \hat{p}_2^{\mu} \, , \qquad
    p_2^{\mu}= \beta_- \hat{p}_1^{\mu}+\beta_+ \hat{p}_2^{\mu} \, ,
\end{align}
with $\beta_{\pm}=(1\pm \beta)/2$, $\beta = \sqrt{1-m^2}$ and $m=2M/\sqrt{s_{12}}$. Therefore, since the momentum of the gluon is labelled in the virtual contribution as $q_1$ and LTD shall set it on-shell, $q_1^2=0$, the momentum mapping proposed for the merging of virtual and real cross-sections is
\begin{align}\label{eq:mapping1}
    p_r^{\prime \mu} &= q_1^{\mu} \, , \nn\\
    p_1^{\prime \mu} &= (1-\alpha_1)\,\hat{p}_1^{\mu} + (1-\gamma_1)\, \hat{p}_2^{\mu} -q_1^{\mu} \, , \nn \\
    p_2^{\prime \mu} &= \alpha_1\,  \hat{p}_1^{\mu}+\gamma_1 \, \hat{p}_2^{\mu}  \, ,
\end{align}
where $\alpha_1$ and $\gamma_1$ are parameters determined by imposing on-shellness. Notice that \Eq{eq:mapping1} fulfills momentum conservation. This mapping will be used to remove some IR singularities of the integral related to $\tilde{\delta}(q_1)$, the remaining IR singularities shall cancel through renormalisation. Now, since there are other singular IR behaviours in the real that can be recognised in the virtual, when $q_2^2=M^2$, the second momentum mapping is proposed to be,
\begin{align}\label{eq:mapping2}
    p_r^{\prime \mu} &= (1-\gamma_2)\,\hat{p}_1^{\mu} + (1-\alpha_2)\, \hat{p}_2^{\mu} -q_2^{\mu} \, , \nn \\
    p_1^{\prime \mu} &= \gamma_2\,  \hat{p}_1^{\mu}+\alpha_2 \, \hat{p}_2^{\mu}  \, ,\nn \\
    p_2^{\prime \mu} &= q_2^{\mu} \, , 
\end{align}
with $\alpha_2$ and $\gamma_2$ parameters which are also determined by the on-shell conditions. In virtue of soft and collinear divergences in the real cross-section being only in this two LTD cuts, we proceed to slice the full real element with the general function,
\begin{align}
    \mathcal{R}_i =   \{ y'_{ir} < \min(y'_{jk}) \} \, , \qquad \sum \mathcal{R}_i = 1 \, ,
\end{align}
which in this particular case are explicitly,
\begin{align}
    \theta(y'_{2r} < y'_{1r} )+ \theta(y'_{1r} < y'_{2r} )=1 \, ,
\end{align}
with $y'_{ir}=2 p_i'\cdot p_r'/s_{12}$. The integration domain is well determined and it is shown in Fig.~\ref{fig:dualregions}.

We remark that this arrangement of the real element phase-space is general and it can be applied to any $1\to 3$ real cross-section that is mapped into the $1\to 2$ virtual cross-section.

Let us now compute the virtual decay rate of $A^* \to q \bar{q}$. The LTD representation is given by,
\begin{align}\label{eq:virtAqq}
    \Gamma_{{\rm V},A}^{(1,{\rm R})} = \frac{1}{2\sqrt{s_{12}}} \sum_{i=1}^3 \int {\rm dPS}^{1\to 2} \, 2{\rm Re} \, \langle \mathcal{M}_A^{(0)} \vert\mathcal{M}_A^{(1,{\rm R})} (\widetilde{\delta}(q_i)) \rangle
\end{align}
with the renormalised one loop amplitude computed as,
\begin{align}
    \vert \mathcal{M}_A^{(1,{\rm R})}\rangle = \vert \mathcal{M}^{(1)}_A\rangle -\vert \mathcal{M}_A^{(1,{\rm UV})}\rangle+\frac{1}{2} (\Delta Z_2^{{\rm IR}}(p_1)+\Delta Z_2^{{\rm IR}}(p_2)) \vert \mathcal{M}_A^{(0)}\rangle \, ,
\end{align}
and $\vert \mathcal{M}_A^{(1,{\rm UV})}\rangle$ is the unintegrated UV counter-term of the one-loop vertex correction. The real element, after the splitting of the phase-space in two domains, is written as,
\begin{align}\label{eq:realAqq}
    \widetilde{\Gamma}_{{\rm R},A, i} ^{(1)} = \frac{1}{2\sqrt{s_{12}}} \int {\rm dPS}^{1\to 3 } \, \vert \mathcal{M}_{A^* \to q\bar{q}g}^{(0)} \vert^2 \, \mathcal{R}_i(y'_{ir}<y'_{jr}) \, , \qquad i,j = \{1,2\} \, .
\end{align}

\begin{figure}[t]
    \centering
    \includegraphics[scale=0.47]{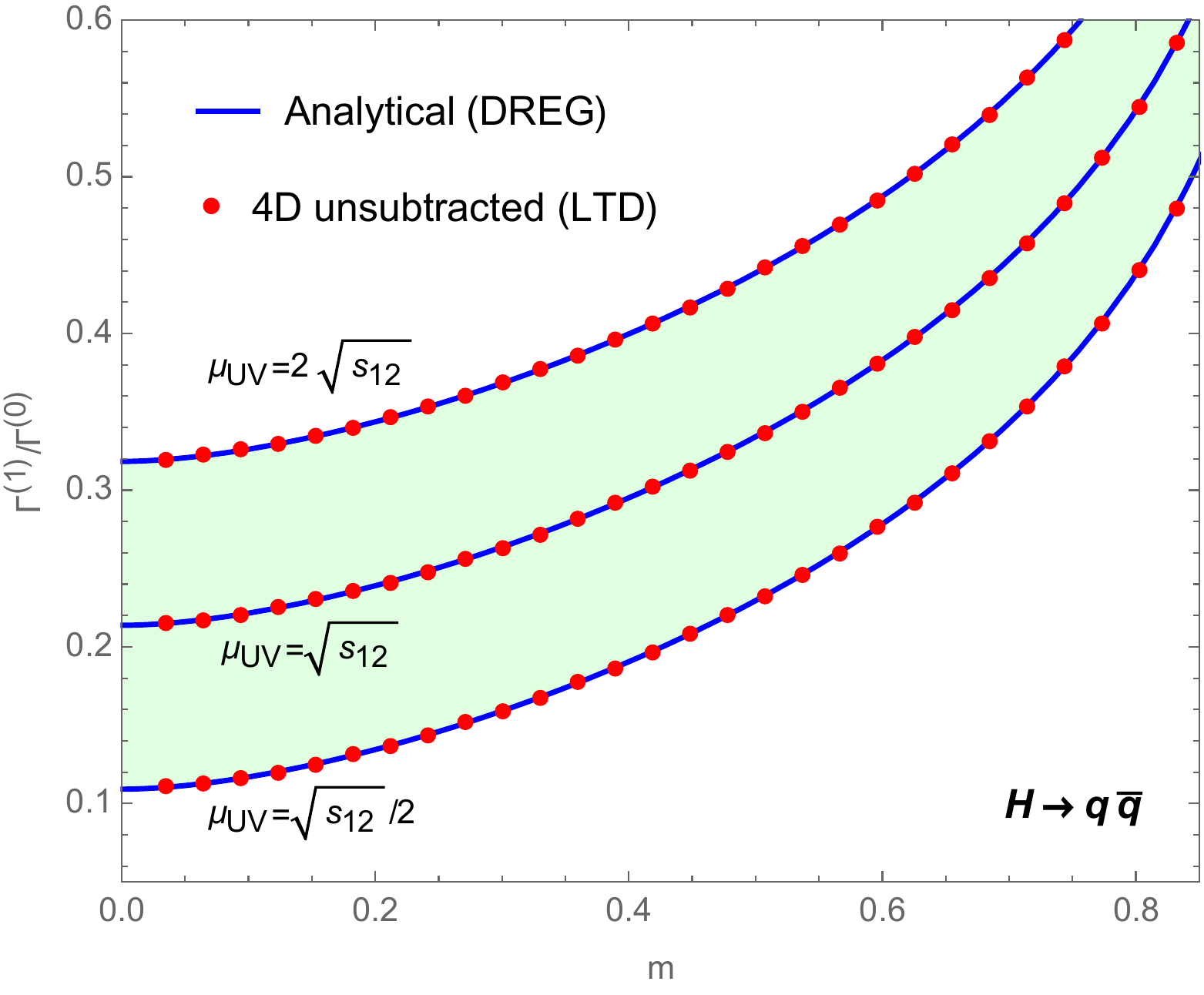}
    \includegraphics[scale=0.47]{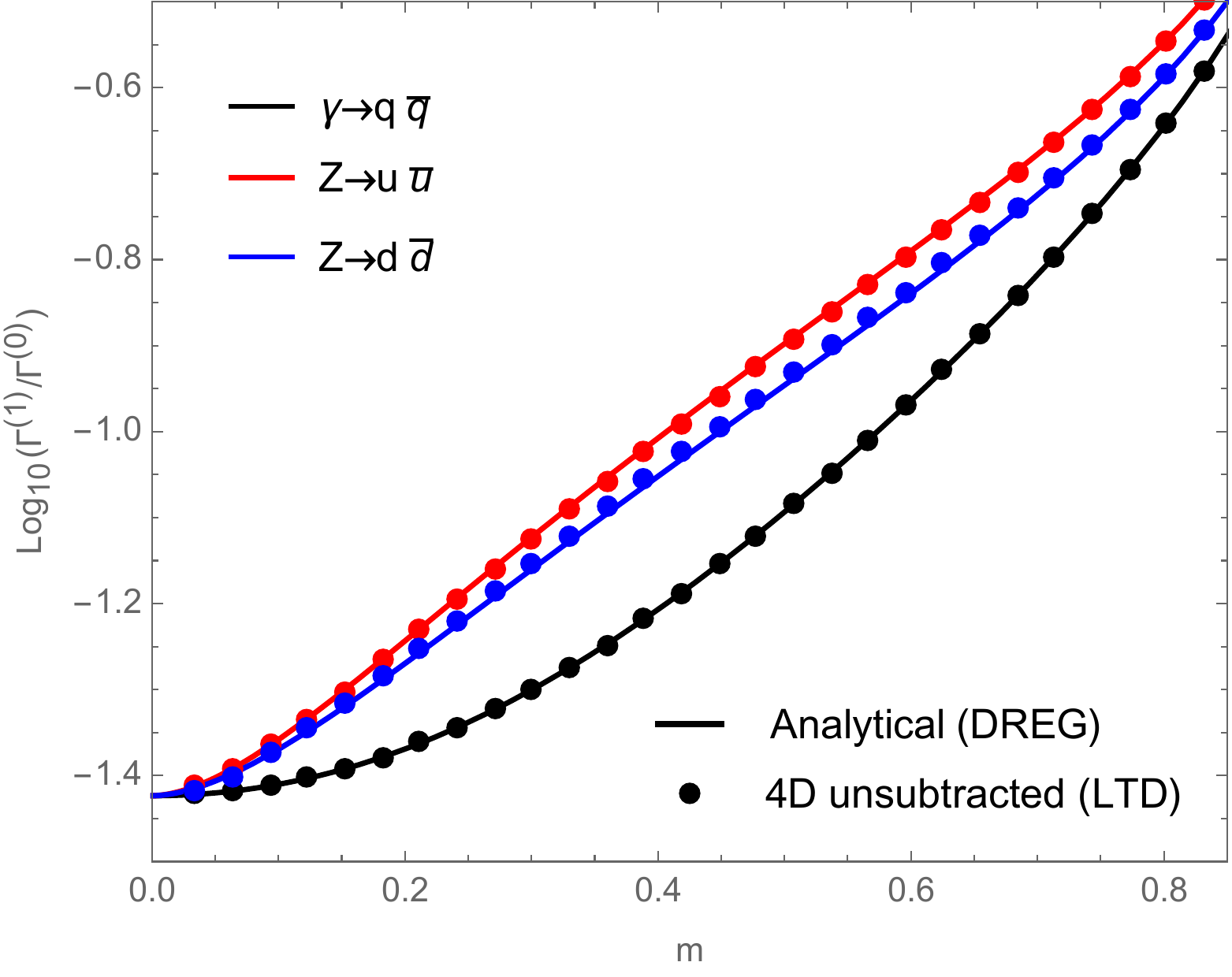}
    \caption{NLO total decay rate for $A^*\to q\bar{q}$ normalised to its corresponding LO decay rate. The solid lines correspond to DREG analytic result while dots were computed numerically in the FDU formalism for the standard Higgs boson (left panel) and off-shell vector bosons (right panel).
    }
    \label{fig:FDUresults}
\end{figure}

The sum of the virtual and real contributions in \Eq{eq:virtAqq} and \Eq{eq:realAqq} is a single integral in the loop-three momentum. It is UV and IR finite locally and can be calculated numerically with $\epsilon = 0$. The results, normalised to the LO decay rate $\Gamma_A^{(0)}$, are presented in Fig.~\ref{fig:FDUresults}. The agreement with the analytical predictions is excellent in all cases. Furthermore, we notice that the massless quark scenario is recovered smoothly in the FDU however, in DREG this is not the case since individual contributions are not smoothly defined in that limit.

%% file: draft_Francesco.tex
In the following, we will focus on the NLO QCD corrections to cross-sections involving massless colored particles and any number of other
colourless SM particles. The extension of the following ideas to the
massive case has been worked out following the same reasoning, which is presented in Ref. \cite{Prisco:2020kyb}. The reduction of any one-loop amplitude to a combination of scalar integrals
is an easy task compared to the two-loop case.
Of course, such a reduction does not alter the structure of the divergences
of the amplitude but to properly reproduce all the IR singularities, in addition
to the wave function renormalisation, one has to keep also the massless
bubble integrals that are formally discarded usually (as already explained in Sec. \ref{sec:FDUinto}). Once all the box functions are expressed in terms of their corresponding
$6-$dimensional version, it turns out that the IR divergences of a
renormalised one-loop amplitude interfered with the leading order are
given by the triangle scalar functions with two external partons connected by a
massless propagator times the corresponding color connected tree level
amplitude multiplied $s_{ij}=2\,p_i \cdot p_j$, the massless bubbles times
(1/2) the Born for every external (gluon) quark external particle multiplied
times the corresponding Casimir, and the external wave function renormalisation
constant time the Born.
The above set of IR singularities is universal and, in fact, it reproduces the known formula for the poles of any one-loop amplitude in QCD. 
Inspired by the LTD theorem, we considered the cuts of these loop
integrals that capture their infrared and collinear divergences. As a result
of this procedure one obtains a novel subtraction scheme. For the sake of simplicity
we will restrict the discussion to the case of a colourless initial state and
so we will consider colored radiation exchanged only among the final state partons.
Our general formula for the dual counter-term in the case of a quark emitting
a gluon is the following:
\begin{equation}
\label{def:count_qg}
\sigma_{qg,b}^{DS} \equiv 8\pi\alpha_S \mathcal{N}_{in} \int d \Phi_m \,_m\langle 1, \dots, m| \mathbf{T}_{ac} \mathbf{T}_b |1, \dots, m \rangle_m  \left[  V_{q_ag_c,b} (p_{i}, p_j) + G_{q_ag_c,b} (p_{i},p_j) \right] \, ,
\end{equation}
where $\mathcal{N}_{in}$ represents all the non-QCD factors, including the symmetry factor for identical partons in the final state, $|1, \dots, m \rangle_m$ is the $m$-parton Born amplitude and $\mathbf{T}_{ac} \mathbf{T}_b$ is a colour-charge operator acting on the colour indices of the partons $ac$ and $b$, respectively.
The functions $V_{qg,b}$ and $G_{qg,b}$ in Eq.\eqref{def:count_qg} are defined by
\begin{align*}
\label{def:count_qg_comp}
V_{qg,b} &\equiv \int_{q_1} \tilde{\delta} (q_1) \,\mathcal{R}_1 \left[- \frac{2 s_{ij}}{(- 2 q_1 \cdot p_i)(2 q_1 \cdot p_j)} +  \frac{2}{(-2 q_1 \cdot p_i)}  \right] \, , \\
G_{qg,b} &\equiv - \frac{(1-\varepsilon)}{s_{ij}} \int_{q_1} \tilde{\delta} (q_1) \,\mathcal{R}_1  \frac{2 q_1 \cdot p_j}{(-2 q_1 \cdot p_i)} \, ,
\numberthis
\end{align*}
where $V_{qg,b}$ groups together the dual contributions from the reduction of the virtual amplitude and $G_{qg,b}$ is the contribution of the quark wave function renormalisation. In the above formulas $p_i$ and $p_j$ represent the momentum of the emitter (with color index $ac$) and spectator (color index $b$) in the Born kinematic respectively, while $q_1$ is the loop momentum that is associated to the gluon in the real radiation kinematic.
In the case a gluon is radiated collinear to another gluon the corresponding dual counter-term $\sigma^{DS}_{gg,b}$ is given by
\begin{equation}
\label{def:count_gg}
\sigma^{DS}_{gg,b}\equiv 8\pi\alpha_S \mathcal{N}_{in} \int \! d \Phi_m \,_m\langle 1, \dots, m| \mathbf{T}_{ac} \mathbf{T}_b  \left[  V^{\mu\nu}_{gg,b} (p_i, p_j) + G^{\mu\nu}_{gg,b} (p_i,p_j) \right]  |1, \dots, m \rangle_m \, ,
\end{equation}
where the symmetrisation $p'_a \leftrightarrow p'_c$ has to be performed on the right-hand side. Once again, in Eq. \eqref{def:count_gg} we have extracted only the dual contributions with $q_1$ on-shell, obtaining
\begin{subequations}
\label{def:count_gg_comp}
\begin{align}
\label{def:count_gg_comp_V}
V^{\mu\nu}_{gg,b} \!\! &\equiv\! - \!\!\int_{q_1} \tilde{\delta} (q_1) \mathcal{R}_1 \left[- \frac{2 s_{ij}}{(- 2 q_1 \cdot p_i) (2 q_1 \cdot p_j)} + \frac{1}{(-2 q_1 \cdot p_i)} \right] g^{\mu \nu} \, ,\\
\label{def:count_gg_comp_G}
G^{\mu\nu}_{gg,b} \!\! &\equiv\! - \!\!\int_{q_1} \!\!  \frac{\tilde{\delta} (q_1) \mathcal{R}_1}{(-2 q_1 \cdot p_i)}  \!\left[ g^{\mu \nu}\! + \frac{d-2}{(-2 q_1 \cdot p_i)}\!\! \left( \frac{q_1 \cdot p_j}{p_i \cdot p_j} -1 \right) \!\!
\left( q_1^{\mu} - \frac{q_1 \cdot p_i}{p_i \cdot p_j} p_j^{\mu} \right)\!\! \left( q_1^{\nu} - \frac{q_1 \cdot p_i}{p_i \cdot p_j} p_j^{\nu} \right) \! \right] \,\,.
\end{align}
\end{subequations}
Finally, the case of a gluon splitting into a collinear quark-antiquark
pair receive contributions only from the wave function contribution so that
the corresponding dual counter-term $\sigma^{DS}_{q\bar{q},b}$ is given by
\begin{equation}
\label{def:count_qq}
\sigma^{DS}_{q\bar{q},b} \equiv 8\pi\alpha_S \mathcal{N}_{in} \int \! d \Phi_m \,_m\langle 1, \dots, m| \mathbf{T}_{ac} \mathbf{T}_b \, G^{\mu\nu}_{q \bar{q},b} (p_i,p_j) |1, \dots, m \rangle_m \, ,
\end{equation}
with
\begin{equation}
\label{def:count_qq_comp}
G^{\mu\nu}_{q\bar{q},b} \!\! \equiv\! \!\!\int_{q_1} \!\!  \frac{\tilde{\delta} (q_1) \mathcal{R}_1 T_R N_f}{C_A (-2 q_1 \cdot p_i)}  \!\left[ g^{\mu \nu}\! \!+\! \frac{4}{(-2 q_1 \cdot p_i)}\!\! \left( \frac{q_1 \cdot p_j}{p_i \cdot p_j} -1 \right) \!\!
\left( q_1^{\mu} - \frac{q_1 \cdot p_i}{p_i \cdot p_j} p_j^{\mu} \right)\!\! \left( q_1^{\nu} - \frac{q_1 \cdot p_i}{p_i \cdot p_j} p_j^{\nu} \right) \! \right].
\end{equation}
Once a proper momentum mapping is considered, these counter-terms can be analytically integrated over the
region in which $p_a \cdot p_c < p_b \cdot p_c $ in the real phase space and added back to the virtual
contribution. This last operation corresponds to extract all the infrared and collinear divergences out
of the virtual matrix element. We have used the mapping adopted in the seminal work by Catani and Seymour~\cite{Catani:1996vz}.

\subsection{\texorpdfstring{$\gamma^* \to 3$}{gamma* -> 3} jets at NLO}
\label{ssec:Gamma3Jets}
As an example of application, we consider the NLO correction to the three-jet production in $e^+e^-$ annihilation $\gamma^* \! \rightarrow q(p_1) \, \bar{q}(p_2) \, g(p_3)$. For simplicity we will restrict to the case of gluon radiation, namely $\gamma^*  \rightarrow q(p'_1)\, \bar{q}(p'_2)\, g(p'_3)\, g(p'_4)$. The case of four quark production can be treated along the same lines.
By inspecting the virtual sector, we find six emitter-spectator pairs for the process $\gamma^* \rightarrow q(p_1) \, \bar{q}(p_2) \, g(p_3)$, that are
$\{ q \bar{q}, qg,\bar{q} g\}$ and the three pairs where the emitter and the spectator switch. Once a real kinematic configuration $(p'_1, p'_2, p'_3, p'_4)$ is generated, the six kinematic invariants $s_{ij}$ are analyzed and as a consequence six dual counter-terms are activated. To each emitter-spectator pair, we associate the following dual counter-terms:
\begin{align*}
\label{scheme:qqgg}
q \bar{q} \quad &\longrightarrow \quad\left\{ \quad
\begin{aligned}
  && g(p'_3)\,\, \text{as radiation:} &&\quad \sigma^{DS}_{13,2} \quad \text{if}\,\, s_{13} < s_{23}, &&\quad \sigma^{DS}  _{23,1} \quad \text{if}\,\, s_{13} > s_{23}\\
&& g(p'_4)\,\, \text{as radiation:} &&\quad \sigma^{DS}_{14,2} \quad\text{if}\,\, s_{14} < s_{24}, &&\quad \sigma^{DS}_{24,1} \quad \text{if}\,\, s_{14} > s_{24}
\end{aligned} \right. \\
q g \quad &\longrightarrow \quad\left\{ \quad
\begin{aligned}
&& g(p'_3)\,\, \text{as radiation:} &&\quad \sigma^{DS}_{13,4} \quad\text{if}\,\, s_{13} < s_{34}, &&\quad \sigma^{DS}_{43,1} \quad \text{if}\,\, s_{13} > s_{34}\\
&& g(p'_4)\,\, \text{as radiation:} &&\quad \sigma^{DS}_{14,3} \quad\text{if}\,\, s_{14} < s_{34}, &&\quad \sigma^{DS}_{34,1}\quad \text{if}\,\, s_{14} > s_{34}
\end{aligned} \right. \\
\bar{q} g \quad &\longrightarrow \quad\left\{ \quad
\begin{aligned}
&& g(p'_3)\,\, \text{as radiation:} &&\quad \sigma^{DS}_{23,4} \quad\text{if}\,\, s_{23} < s_{34}, &&\quad \sigma^{DS}_{43,2} \quad \text{if}\,\, s_{24} > s_{34}\\
&& g(p'_4)\,\, \text{as radiation:} &&\quad \sigma^{DS}_{24,3} \quad \text{if}\,\, s_{24} < s_{34}, &&\quad \sigma^{DS}_{34,2} \quad \text{if}\,\, s_{24} > s_{34}
\end{aligned} \right. \numberthis
\end{align*}
Since the process involves only three colored partons, the colour algebra factorises and one has $\mathbf{T}_1 \mathbf{T}_2 = C_A/2- C_F$ and $\mathbf{T}_1 \mathbf{T}_3 = \mathbf{T}_2 \mathbf{T}_3 = - C_A/2$.
In the virtual sector, we use the integrated version of the dual counter-terms that are given by
\begin{align*}
\sigma^{DS}_{q,\bar{q}} &\sim  \frac{\alpha_S}{2\pi} \frac{(4 \pi)^{\varepsilon}}{\Gamma (1 - \varepsilon)}  \left( \frac{C_A}{2} - C_F \right) \int d \Phi_2 \left( \frac{\mu^2}{s_{12}} \right)^{\varepsilon} | A^{(0)}_{q \bar{q} g} |^2 \left[ \frac{1}{\varepsilon^2} + \frac{3}{2\varepsilon} + 3 + 4 \log(2) -\frac{\pi^2}{2}  \right] \\
\sigma^{DS}_{\bar{q}, q}&\sim  \frac{\alpha_S}{2\pi} \frac{(4 \pi)^{\varepsilon}}{\Gamma (1 - \varepsilon)}  \left( \frac{C_A}{2} - C_F \right) \int d \Phi_2 \left( \frac{\mu^2}{s_{12}} \right)^{\varepsilon} | A^{(0)}_{q \bar{q} g} |^2 \left[ \frac{1}{\varepsilon^2} + \frac{3}{2\varepsilon} + 3 + 4 \log(2) -\frac{\pi^2}{2}  \right] \\
\sigma^{DS}_{q, g} &\sim - \frac{\alpha_S}{2\pi} \frac{(4 \pi)^{\varepsilon}}{\Gamma (1 - \varepsilon)} \frac{C_A}{2} \int d \Phi_2 \left( \frac{\mu^2}{s_{13}} \right)^{\varepsilon} | A^{(0)}_{q \bar{q} g} |^2 \left[ \frac{1}{\varepsilon^2} + \frac{3}{2\varepsilon} + 3 + 4 \log(2) -\frac{\pi^2}{2}  \right] \\
\sigma^{DS}_{g,q} &\sim - \frac{\alpha_S}{2\pi} \frac{(4 \pi)^{\varepsilon}}{\Gamma (1 - \varepsilon)} \frac{C_A}{2}  \int d \Phi_2 \left( \frac{\mu^2}{s_{13}} \right)^{\varepsilon} | A^{(0)}_{q \bar{q} g} |^2 \left[ \frac{1}{\varepsilon^2} + \frac{11}{6 \varepsilon} + \frac{55}{18} + \frac{14}{3} \log(2) -\frac{\pi^2}{2} \right] \\
\sigma^{DS}_{\bar{q},g} &\sim - \frac{\alpha_S}{2\pi} \frac{(4 \pi)^{\varepsilon}}{\Gamma (1 - \varepsilon)} \frac{C_A}{2}  \int d \Phi_2 \left( \frac{\mu^2}{s_{23}} \right)^{\varepsilon} | A^{(0)}_{q \bar{q} g} |^2 \left[ \frac{1}{\varepsilon^2} + \frac{3}{2\varepsilon} + 3 + 4 \log(2) -\frac{\pi^2}{2}  \right] \\
\sigma^{DS}_{g, \bar{q}} &\sim - \frac{\alpha_S}{2\pi} \frac{(4 \pi)^{\varepsilon}}{\Gamma (1 - \varepsilon)} \frac{C_A}{2}  \int d \Phi_2 \left( \frac{\mu^2}{s_{23}} \right)^{\varepsilon} | A^{(0)}_{q \bar{q} g} |^2 \left[ \frac{1}{\varepsilon^2} + \frac{11}{6 \varepsilon} + \frac{55}{18} + \frac{14}{3} \log(2) -\frac{\pi^2}{2}  \right] \numberthis
\end{align*}
where $| A^{(0)}_{q \bar{q} g} |^2$ is the Born amplitude for the process $\gamma^* \rightarrow q \bar{q} g$ and we have neglected terms of order $\mathcal{O}(\varepsilon)$.

\begin{figure}[t]
	\centering
	\includegraphics[width=0.7\linewidth]{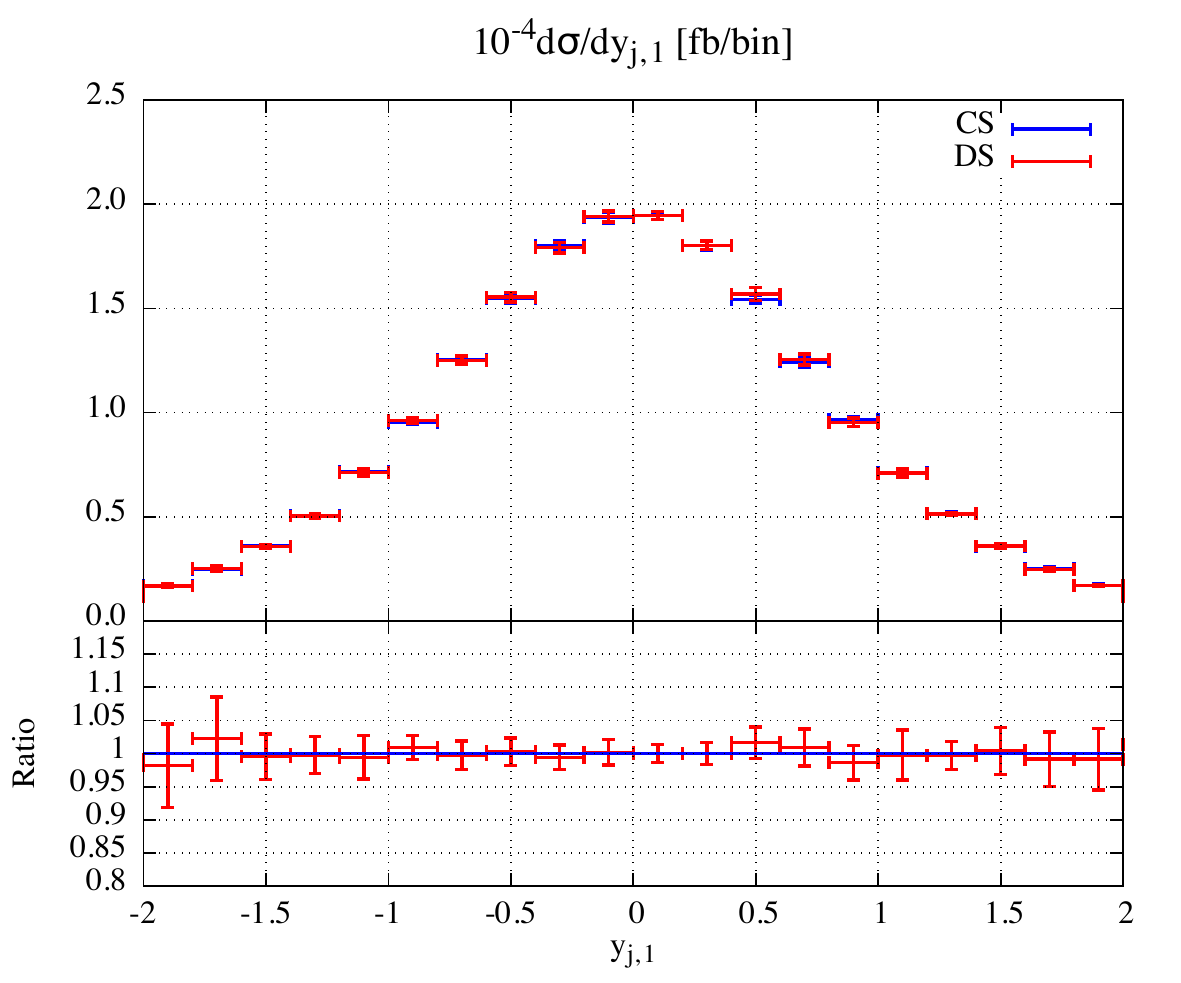}
	\caption{Rapidity distribution of the most energetic jet in $e^+e^- \to \gamma^* \to 3$ jets at $\sqrt{s} = 125\, \text{GeV}$.
         Jets are clustered using the Durham algorithm with $ y_\text{cut} = 0.05$
         Only the NLO corrections are plotted. Results obtained using Dual Subtraction (DS) results are shown in red,
         while the ones obtained using Catani-Seymour (CS) dipoles are shown in blue.
	 The error bars correspond to the statistical error given by the Monte Carlo integration.}
	\label{fig:epem3j_cor_j1_rap}
	\end{figure}
	
In Fig. \ref{fig:epem3j_cor_j1_rap} we show the differential rapidity distribution of the most energetic jet in $e+e-$ annihilation at $\sqrt{s} = 125\, \text{GeV}$. Jets are clustered using the Durham jet algorithm with resolution parameter set at $y_\text{cut}=0.05$. A perfect agreement with the same computation performed using Catani-Seymour dipoles is observed. The same level of agreement is observed for any other differential distribution.

In this way, we have shown that a fully local implementation of an LTD-inspired method for computing cross-sections is reliable. These studies complement the ones presented in Sec. \ref{sec:FDUinto}, indicating that the FDU framework constitutes a powerful strategy for reaching higher-order corrections with fully numerical methods. An extension of this technology to calculate N$^2$LO corrections for IR-safe observables is being actively studied.